\documentclass[twocolumn,aps,prb]{revtex4-2}
\usepackage{amssymb}
\usepackage{graphicx}
\usepackage{hyperref}
\usepackage{xcolor}
\usepackage{multirow}
\usepackage{amsmath}
\begin{document}

\preprint{APS/123-QED}

%\title{Electric field induced spin current generation in gapped Rashba systems}
\title{
Role of Berry curvature in the generation of spin currents in Rashba systems}

\author{Priyadarshini Kapri}
\email{pkapri@iitk.ac.in}
\author{Bashab Dey}
\author{Tarun Kanti Ghosh}

\affiliation{Department of Physics, Indian Institute of Technology-Kanpur, 
Kanpur-208 016, India }

\date{\today}

\begin{abstract}
We study the background (equilibrium), linear and nonlinear spin currents in 2D Rashba spin-orbit coupled systems with Zeeman splitting and in 3D noncentrosymmetric metals using modified spin current operator by inclusion of the anomalous velocity.
%semi-classical Boltzmann transport formalism. 
%Our calculation involves energy dependent relaxation time obtained by solving the Boltzmann transport equations with interband, intraband/interbranch and intrabanch scattering self-consistently in presence of  the short-range electron impurity. 
The linear spin Hall current arises due to the anomalous velocity of charge carriers induced by the Berry curvature. The nonlinear spin current occurs due to the band velocity  and/or the anomalous velocity. %The nonlinear spin current may occur due to the anomalous velocity.
For 2D Rashba systems, the background spin current saturates at high Fermi energy (independent of the Zeeman coupling), linear spin current exhibits a plateau at the `Zeeman' gap and nonlinear spin currents are peaked at the gap edges. The magnitude of the nonlinear spin current peaks enhances with the strength of Zeeman interaction. The linear spin current is polarized out of plane, while the nonlinear ones are polarized in-plane. We witness pure anomalous nonlinear spin current with spin polarization along the direction of propagation. In 3D noncentrosymmetric metals, background and linear spin currents are monotonically increasing functions of Fermi energy, while nonlinear spin currents vary non-monotonically as a function of Fermi energy and are independent of the Berry curvature. These findings may provide useful information to manipulate spin currents in Rashba spin-orbit coupled systems.
\end{abstract}

\maketitle

\section{Introduction}   
Spintronics is a field where the spin and
charge degrees of freedom  of the carriers are used for controlling the properties of materials and devices \cite{Wolf,Zutic,Bader,Schliemann}. Thus, the generation, manipulation, and detection of spin have received enormous impetus %because they are the leading aspects%
in the field of spintronics.
%In semiconductor spintronics, the control over electron's  spin is the key ingredient for information processing.  currents and hold a promise to combine
%storage, memory and computing functionalities.   Hence, the generation and control of spin current .    
It has been a substantial issue to uncover more efficient ways to 
generate the spin current. Various techniques are available for the generation of spin current, such as, the spin injection or pumping from proximity ferromagnets \cite{Datta,Gardelis,Schmidt,Hu,Tombros,Xiao}, spin battery \cite{Saitoh,Ando,Dushenko,Lesne,Kondou}, optical injection methods that depend
on optical selection rules \cite{Ganichev,Stevens} etc. 

Recently, focus has been paid on the generation of spin current and their manipulation without using any magnets, where the spin-orbit (SO) coupling plays a crucial
role. The spin-orbit interaction
is the coupling between the spin and momentum, which intrinsically occurs in all materials, due to relativistic effects. However, lack of surface inversion symmetry
in the confinement potential of electrons in a quantum well or
a heterostructure gives rise to a particular type of SO interaction known as Rashba spin-orbit interaction (RSOI) \cite{Rashba,Bychkov}. The RSOI has great importance in the emerging
field of spintronics for fabricating novel devices with the possibility of being able to tune the RSOI
strength by an external gate voltage or other techniques \cite{Nitta,Engels}. 
%Besides, lack of crystal inversion symmetry, where an electron
%feels an asymmetry in the crystal potential due to the lack of bulk-inversion %symmetry, may cause to another type of SO interaction known as 
%Dresselhaus spin-orbit interaction \cite{Dresselhaus}. 
%Spin-orbit coupled systems provide a compelling probe for the investigations
%fascinated on bringing up phenomena which can show the way to initiate a spin-dependent transport paradigm for futuristic spintronic devices \cite{,Zutic}.

 In Ref. \cite{Rashba2,Rashba3}, Emmanuel I. Rashba showed that a finite spin current exists in noncentrosymmetric systems under thermodynamical equilibrium (i.e. in absence of an electric field), which is known
as background equilibrium spin current and stated that such background equilibrium spin current can not
transport and accumulate electron spins. This
is considered to be the byproduct of using the conventional
definition of spin current operator in a spin non-conserving system. So, the modification of conventional definition of spin current operator was proposed to eliminate such equilibrium  spin current.  Subsequently, immense debate
regarding the definition of the spin current had started \cite{Shi,Sun1,Wang1,Wang2}. However, in \cite{Sun2}, the authors gave physical arguments to show that such equilibrium spin current in spin-orbit coupled system is the persistent spin current. It was asserted that
spin-orbit interaction plays the role of the spin driving
force which leads  to a pure persistent spin current. Moreover, they argued that the conventional definition of the spin current does not need to be modified, as the equilibrium spin current, the non conservation of spin current, and the violation of the Onsager relation are intrinsic properties of spin
transport irrespective of the definition of the spin current operator. There have been many
works on the topic of persistent spin current \cite{Dolcini1,Dolcini2,Loss,Splettstoesser,Schutz,Usaj}.  Further, this persistent spin
current can also generate an electric field \cite{Schutz,Sun3} which offers a way for its detection. In Ref.
\cite{Sonin}, the author has made an interesting proposal to detect such equilibrium spin current by studying induced
mechanical torques on a cantilever at the edges of the
Rashba system.  Further, in  Ref. \cite{Newman} the authors have demonstrated how to detect DC spin current with a static field applied at different orientations within the plane of the sample through using an epitaxial antiferromagnetic NiO layer.

In a spin-orbit coupled system, an electrical charge current  
can yield a transverse pure spin current with polarization perpendicular to the plane of the charge and spin current. This is known as spin Hall current which arises mainly due to an intrinsic mechanism governed by the geometry of the Bloch wave functions \cite{Murakami1,Murakami3,Sinova,Wunderlich,Paul}. Further, it may appear  %aslso due the the  Spin-orbit coupled systems produce a 
because of the extrinsic mechanism such as the skew scattering \cite{Hirsch,Zhang,Kato1}.  
Among several other possibilities, the spin Hall effect (SHE) for creating and manipulating the spin current has gained
its distinct place \cite{Kato1,Day}.
% 
%electric field generated by the spin current 
%
%polarized light beam and second-order nonlinear optical effect.
%Based on symmetry arguments, it is shown that longitudinal spin current which is proportiocal to the external electric field vanishes identically. The first non-vanishing longitudinal spin current is quadratic in the external electric field.

Discrete symmetries of the Hamiltonian viz. inversion symmetry (IS) and time reversal symmetry (TRS) play a crucial role in determining the fate of spin current. It has been shown that the presence of IS and TRS requires even and odd order contributions of electric field to the spin current to vanish, respectively \cite{Hamamoto}.  Thus,  breaking atleast one of the symmetries is a necessary (but not sufficient) condition to produce finite spin current.  
In recent years there is a growing interest on the generation of nonlinear spin current \cite{Hamamoto,Pan,Hongyi} in spin-orbit coupled systems. 
%In ,Pan,Hongyi}, the authors have reported the generation 
The nonlinear spin can arise
in a 2D crystal of Fermi surface anisotropy \cite{Hongyi} or in a noncentrosymmetric spin-orbit coupled system \cite{Hamamoto,Pan} with a simple application of an electric field  {\bf E}. 
%{\color{green}  Studies in Ref. \cite{Hamamoto,Pan,Hongyi} show the generation of DC spin current by an AC electric
%field and assure the controllability of polarization of spin current by adjusting the direction of the applied electric field}. 

Motivated by the above discussion, we redefine the spin current operator by 
inclusion of the anomalous velocity so that it can give rise to both the linear spin Hall current as well as the nonlinear spin current. We provide a systematic study 
of spin Hall current along with the nonlinear spin current in Rashba systems having different Fermi surface topology below and above the band touching point (BTP).
It is explicitly shown that the spin Hall current arises solely due to the anomalous velocity. We also find that the nonlinear spin current may arise due to the anomalous velocity. 
In the study of nonlinear spin current, we 
consider energy-dependent relaxation time by solving the Boltzmann transport 
equations self-consistently.

This paper is organized as follows. In Sec. \ref{Sec2}, we provide a discussion on the formalism of spin current for a generic two band system. Section \ref{Sec3} includes the basic information of 2D gapped Rashba system and its corresponding results on spin currents. In Sec. \ref{Sec4}, we present the general information as well as the results on 3D Rashba system. 
%In Sec. V and Sec. VI, we present the information of 3D Rashba system and their corresponding results respectively. 
Finally, we conclude and summarize our main results in Sec. \ref{Sec5}.

\section{Formalism of spin current}
\label{Sec2}
Here we provide a general formalism of spin current 
for a generic two-band system in presence of an external electric
field. First we describe the ground state properties of a generic
two-band system. Then we discuss the modified 
Fermi-Dirac distribution function due to an applied electric field.
Finally we present a general expression of spin current in terms 
of the density of states and energy dependent scattering time. 

\subsection{Generalized system}
\label{seca} 
A generic Hamiltonian of a two-band system is expressed in the form
\begin{equation}
\label{eqH}
H({\bf k})=\frac{{ \hbar^2 \bf k^2}}{2m^{*}}\sigma_0+{\boldsymbol \sigma}\cdot \boldsymbol d({\bf k)},
\end{equation}
where $m^{*}$ is the effective mass of a charge carrier,  
$\sigma_0$ is the $2 \times 2$ identity matrix,  
$\sigma_{x,y,z}$ are the Pauli's spin matrices, and 
${\boldsymbol d({\bf k})} = \{d_x({\bf k}), d_y({\bf k}), d_z({\bf k})\}$ 
with ${\bf k}$ being the wavevector of the charge carrier. 
The energy spectra of the system is obtained as
\begin{equation}
\epsilon_{\lambda}({\bf k})=\frac{{ \hbar^2 k^2}}{2m^{*}} + \lambda d({\bf k}),
\end{equation}
with $\lambda=\pm$ denoting the band indices and 
$d({\bf k})=\sqrt{d_x^2({\bf k}) + d_y^2({\bf k}) + d_z^2({\bf k})} $. 
In general, there are two spin-split Fermi surfaces due to the presence of 
the kinetic energy term in Eq. (\ref{eqH}), as compared to a single
Fermi surface for massless case.
The corresponding eigenstates are
\begin{eqnarray} \label{gen-wf}
|{\bf k}, +\rangle =\left(\begin{array}{c}
\cos\frac{{\theta^{\prime}}}{2}e^{-i \phi^{\prime}} \\
\sin\frac{{\theta^{\prime}}}{2}  \end{array} \right) \hspace{0.01in}; 
\hspace{0.01in}
|{\bf k}, -\rangle =\left(\begin{array}{c}
\sin\frac{{\theta^{\prime}}}{2}e^{-i \phi^{\prime}} \\
-\cos\frac{{\theta^{\prime}}}{2}  \end{array} \right),\\\nonumber
\end{eqnarray}
where $\cos\theta^{\prime}=d_z({\bf k})/d({\bf k})$ and  $\tan\phi^{\prime}=d_y({\bf k})/d_x({\bf k})$.

The spin orientation of a charge carrier with wave vector ${\bf k}$ at the
band $\lambda $ is given by
$\langle\boldsymbol \sigma\rangle_{\lambda} = 
\lambda \boldsymbol d({\bf k)}/d({\bf k})$
and thus $\langle\boldsymbol \sigma\rangle_{\lambda}\cdot {\bf k}
= \lambda \boldsymbol d({\bf k)}\cdot {\bf k} /d({\bf k})$.
The Berry curvature of a given band can be obtained from
the following expression: 
%\cite{Xiao2}
${\bf \Omega_{\lambda}(k)} = i 
{\bf \nabla_{{\bf k}}} \times {\bf \langle {\bf k}, \lambda| 
\nabla_{{\bf k}}|{\bf k},\lambda\rangle}$.
The band velocity of a charge carrier is
${\bf v}_{b}^{\lambda} = (1/\hbar) 
\nabla_{\bf k}\epsilon_{\lambda}({\bf k})$.
In presence of an external electric field ${\bf E}$, a
charge carrier with charge $q= -e$ acquires an additional velocity 
(transverse to the electric field direction) 
${\bf v}_{a}^{\lambda} = (e/\hbar) 
{\bf \Omega }_{\lambda}({\bf k}) \times {\bf E}$.
This additional velocity is also termed as anomalous velocity.
Thus, there may be a transverse current to the electric field direction 
for a system having non-zero Berry curvature. 
Therefore, the generalized velocity expression can be written as 
$$
{\bf v}^{\lambda} = \frac{1}{\hbar} \nabla_{\bf k}\epsilon_{\lambda}({\bf k})+\frac{e}{\hbar}\bf{\Omega_{\lambda}(k)\times E}.
$$

\subsection{The approximation in carriers' distribution function}
\label{secb}
When the system is subjected to a local perturbation induced 
by a spatially uniform electric field ${\bf E}$,
the electron energy is modified to 
$\epsilon^\prime({\bf r}, {\bf k}) = \epsilon({\bf k}) + e {\bf E \cdot r}$, 
where ${\bf r}$ is the spatial coordinate. 
%(for the time being we have dropped the band index $\lambda$).
Typically this change in energy is very weak as compared to
the Fermi energy $\epsilon_F$. 
The Fermi-Dirac distribution function 
$f({\bf r},{\bf k}) = 
(1 + e^{\beta [\epsilon^\prime({\bf r}, {\bf k})-\epsilon_F]} )^{-1}$
with $\beta = 1/(k_BT)$ can be expanded in a series of terms proportional 
to powers of the electric field ${\bf E}$. The linear term is bound 
to reproduce the solution of the Boltzmann transport equation (BTE) and 
thus the spatial coordinate must be in the form  
${\bf r}= {\bf v}_{b} \tau(\epsilon)$ with $\tau(\epsilon)$ being 
the energy-dependent relaxation time.
With this consideration, the modified distribution function in presence 
of the external electric field becomes 
\cite{Pan,Gao}
\begin{equation}
\label{eqdfn}
f(\epsilon, {\bf E}, \tau) = 
\sum_{n}f_{n} = \sum_n \frac{[e\tau(\epsilon) {\bf E} \cdot {\bf v}_{b}]^n}{n!} 
\frac{\partial^n f_0(\epsilon)}{\partial \epsilon^n}.
\end{equation}
Here $n=0,1,2...$ and  $f_{0}(\epsilon) = 
(1 + e^{\beta [\epsilon({\bf k})-\epsilon_F]})^{-1}$ is the equilibrium 
distribution function in absence of the electric field. 
Moreover, $ f_{n}(\epsilon) \sim E^n$ is the $n$-th order deviation   
from the equilibrium distribution function $f_{0}(\epsilon)$ due to 
the applied electric field.

\subsection{Modified definition of spin current}
\label{secc}
The conventional definition of spin current operator is given by
$\hat{{v}}_{b,ij} = (\hat{v}_{b,i} \sigma_{j} + 
\sigma_{j} \hat{v}_{b,i} )/2$,
where the first index $i$ and the second index $j$ indicate the direction of propagation and spin orientation of a charge carrier, respectively, with 
$\hat{v}_{b,i} = \frac{1}{\hbar} 
\frac{\partial H({\bf k}) }{\partial k_{i}} $  being  the 
band velocity operator in $i$ direction.
In this conventional definition of spin current operator, only the band
velocity contribution has been considered, while 
the contribution from the anomalous velocity is completely neglected. The definition of velocity operator including the anomalous term ($\hat{v_i}=\hat{v}_{b,i}+\hat{v}_{a,i}$, where $\hat{v}_{b,i} = \frac{1}{\hbar} 
\frac{\partial H({\bf k}) }{\partial k_{i}} $ and $\hat{v}_{a,i} = -(e/\hbar) \epsilon_{ijk} E_j \Omega_k\sigma_0$) is well known and has been used extensively in literature \cite{Chang,Sundaram,Sodemann}, where the anomalous term is responsible for the well known anomalous Hall effect (AHE). In our study,  similar approach has been used to define the spin current operator to check whether the anomalous part of the spin current operator can produce the spin Hall effect.  Thus, the  redefined  
 spin current operator is given by
\begin{equation}  
\hat{{v}}_{ij} = \hat{{v}}_{b,ij} + \hat{v}_{a,ij}, 
\end{equation} 
where $\hat{v}_{a,ij} = v_{a,i} \sigma_j$ with
$v_{a,i} = -(e/\hbar) \epsilon_{ijk} E_j \Omega_k$  being the
anomalous velocity in $i$ direction.
%The first and second terms of the generalized velocity operator yield the band %velocity ${\bf v}_b$ and the anomalous velocity ${\bf v}_a$, respectively.
%Usually, the anomalous velocity term in the definition of spin current
%is neglected since the Berry curvature vanishes for most of the systems.
%However, we keep the anomalous term in the spin current expression.
Later it will be revealed that the anomalous velocity is solely responsible
for the linear spin Hall current  and may contribute to non-linear spin current.
%Therefore, the spin current operator can be written as
%$ \hat{{v}}_{ij} =  \hat{{v}}_{b,ij} + \hat{{v}}_{a,ij}$, where
%$\hat{{v}}_{b,ij}$ is due to the band velocity and $\hat{{v}}_{a,ij}$
%is due to the anomalous velocity.

The total spin current is given in the form of the integral of the average of 
the generalized spin current operator $\hat v_{ij}$, weighted by the distribution 
function $f(\epsilon, {\bf E})$:
%at a temperature $T$: 
\begin{equation}
\mathcal{J}_{ij}^{(n)} = \frac{\hbar}{2 }  
\sum_{n,\lambda}
\int \frac{ d^D{\bf k} }{(2\pi)^D} 
\langle \lambda, {\bf k}|\hat{{v}}_{ij}| \lambda, {\bf k}\rangle f_{n}. 
\end{equation}
Here $D$ denotes the spatial dimension of the system under consideration.
The spin current of the $n$-th order appearing from the band velocity is given by
\begin{equation}
\label{eqjb}
\mathcal{J}_{b,ij}^{(n)} = \frac{\hbar}{2 } 
\sum_{\lambda} \int \frac{d^D{\bf k}}{(2\pi)^D} 
\langle \lambda, {\bf k}|\hat{{v}}_{b,ij}| \lambda, {\bf k}\rangle f_{n}.
\end{equation}
Similarly, the $(n+1)$-th order spin current appearing from 
the anomalous velocity is given by
\begin{equation}
\label{eqja}
\mathcal{J}_{a,ij}^{(n+1)} = \frac{\hbar}{2 } \sum_{\lambda} 
\int \frac{d^D{\bf k}}{(2\pi)^D} 
\langle \lambda, {\bf k}|\hat{{v}}_{a,ij}| \lambda, {\bf k}\rangle f_{n}.
\end{equation}
Since the anomalous velocity ${\bf v}_a$  $\propto {\bf E}$ due 
to non-zero Berry curvature, the lowest order spin current arising 
from the anomalous velocity is one.  Thus, the total spin current of 
order $n$ is given by $\mathcal{J}_{ij}^{(n)}=\mathcal{J}_{b,ij}^{(n)}+\mathcal{J}_{a,ij}^{(n)}$.

%It should be noted that lacking of inversion symmetry results in 
%zeroth-order spin current $\mathcal{J}_{ij}^{{(0)}}$. 
It is useful to express the various order spin currents in 
terms of the density of states and relaxation time and  thus they (up to second-order) are presented below.
%Now we shall present general expressions of various order of 
%the spin current (up to second-order/non-linear) for multi-band systems. 

The zeroth-order spin current 
($\mathcal{J}_{ij}^{{(0)}} = \mathcal{J}_{b,ij}^{{(0)}}$) 
appears from the band velocity  and hence has the form \cite{Rashba2,Rashba3}
%$T=0$ as$
\begin{equation} 
\mathcal{J}_{ij}^{{(0)}} =  \frac{\hbar}{2^D \pi} 
\sum_{\lambda} \int_{-\infty}^{\infty} d\epsilon \int d\Omega_s  D_\lambda(\epsilon)
\langle \hat{v}_{b,ij} \rangle_{\lambda}f_0,
\end{equation}
where $\Omega_s$ being the solid angle for $3D$ and polar angle for $2D$ systems and $D_{\lambda}(\epsilon) = \frac{1}{(2\pi)^D}
\int d^Dk \delta(\epsilon - \epsilon_{\lambda}({\bf k}))$ is density 
of states (DOS).  
The non-zero value of zeroth-order spin current indicates that 
the spin current persists even in thermodynamic equilibrium 
(i.e. in the absence of an external field). This is not associated 
with the real spin transport and cannot yield any spin injection 
or spin accumulation.  This is  known as the 
background or equilibrium spin current. 

The linear spin current due to the band velocity and driven by an electric field $E_{\eta}$ ($\eta=x,y,z$)  is given as
\begin{equation}
\label{eqja2}
\mathcal{J}_{b,ij}^{^{(1),\eta}}=\frac{(e E_{\eta})\hbar}{2^{D}\pi}\sum_{\lambda}
\int_{-\infty}^{\infty}d\epsilon\mathcal{P_{\lambda}}(\epsilon)
\Big(\frac{\partial f_{0}}{\partial \epsilon}\Big),
\end{equation}
where $\mathcal{P_{\lambda}}(\epsilon)=\int d\Omega_s\tau_{\lambda}(\epsilon)D_{\lambda}(\epsilon)\langle\hat{v}_{b,ij}\rangle_{\lambda}\langle \hat{v}_{b,\eta}\rangle_{\lambda}$
with $\tau_\lambda(\epsilon) $ being the energy-dependent relaxation time.

%[{\bf Can we argue from symmetry considerations that the linear spincurrent due to the band velocity is always zero?}\\

On the other hand, the linear spin current  due to the anomalous velocity has 
the following form
\begin{equation}
\label{eqja1}
\mathcal{J}_{a,ij}^{^{(1),\eta}}=\frac{\hbar}{2^{D}\pi}\sum_{\lambda}
\int_{-\infty}^{\infty}d\epsilon\int d\Omega_s D_{\lambda}(\epsilon)\langle\hat{v}_{a,ij}^{\eta}\rangle_{\lambda} f_{0},
\end{equation} 
with $\eta$ being the direction of electric field and always propagates in the perpendicular direction to the applied 
electric field. Later it will be shown that this linear spin current 
is responsible for the spin Hall effect.
 
The general expression of quadratic spin current
$\mathcal{J}_{b,ij}^{^{(2),\eta}}$ appearing from the band velocity
is given by
\begin{equation}
\label{eqjb2}
 \mathcal{J}_{b,ij}^{^{(2),\eta}}=\frac{(e E_{\eta})^2\hbar}{2^{D+1}\pi}\sum_{\lambda}
\int_{-\infty}^{\infty}d\epsilon\mathcal{H_{\lambda}}(\epsilon)
\Big(\frac{\partial^2 f_{0}}{\partial \epsilon^2}\Big),
\end{equation}
where $\mathcal{H_{\lambda}}(\epsilon)=\int d\Omega_s\tau_{\lambda}^2(\epsilon)D_{\lambda}(\epsilon)\langle\hat{v}_{b,ij}\rangle_{\lambda}\langle \hat{v}_{b,\eta}\rangle ^2_{\lambda}$.
Thus, for an isotropic system, $\mathcal{H_{\lambda}}(\epsilon)=\tau_{\lambda}^2(\epsilon)D_{\lambda}(\epsilon)\int d\Omega_s\langle\hat{v}_{b,ij}\rangle_{\lambda}\langle \hat{v}_{b,\eta}\rangle ^2_{\lambda}$. Performing integration by parts on  Eq. (\ref{eqjb2}), 
the general expression for quadratic spin current arising from the band velocity 
at zero temperature is obtained as
\begin{equation}
\label{eqjb22}
\mathcal{J}_{b,ij}^{^{(2),\eta}}=\frac{(e E_{\eta})^2\hbar}{2^{D+1}\pi} 
\sum_{\lambda}  \mathcal{G_{\lambda}}(\epsilon_F),
%+\frac{\pi^2(K_BT)^2}{6}\frac{d^2\mathcal{G_{\lambda}(\epsilon)}}{d\epsilon^2} 
%\Big|_{\epsilon=\epsilon_F}\Big],
\end{equation}
where $\mathcal{G_{\lambda}}(\epsilon)=\frac{d\mathcal{H_{\lambda}(\epsilon)}}{d\epsilon}$.   
Therefore, zero temperature quadratic spin current $\mathcal{J}_{b,ij}^{^{(2),\eta}}$ depends on the first derivative of density of states (DOS).

Now, the general expression of
quadratic spin current  $\mathcal{J}_{a,ij}^{^{(2),\eta}}$ appearing from the anomalous velocity and driven by an electric field $E_{\eta}$ is given by
%\begin{widetext}
\begin{equation}
\label{eqja2}
\mathcal{J}_{a,ij}^{^{(2),\eta}}=\frac{(e E_{\eta})\hbar}{2^{D}\pi}\sum_{\lambda}
\int_{-\infty}^{\infty}d\epsilon\mathcal{F_{\lambda}}(\epsilon)
\Big(\frac{\partial f_{0}}{\partial \epsilon}\Big),
\end{equation}
where $\mathcal{F_{\lambda}}(\epsilon)=\int d\Omega_s\tau_{\lambda}(\epsilon)D_{\lambda}(\epsilon)\langle\hat{v}_{a,ij}\rangle_{\lambda}\langle \hat{v}_{b,\eta}\rangle_{\lambda}$ with $\langle\hat{v}_{a,ij}\rangle_{\lambda}\propto  {\bf E}$. Hence, at zero temperature the 
$\mathcal{J}_{a,ij}^{^{(2),\eta}}$ has the form $\mathcal{J}_{a,ij}^{^{(2),\eta}}=-\frac{(e E_{\eta})\hbar}{2^{D}\pi}\sum_{\lambda}\mathcal{F_{\lambda}}(\epsilon_F)$. 

 Here we would like to mention that in Ref. \cite{Hamamoto,Pan}, the study of nonlinear spin currents in 2D Rashba systems are carried out with constant relaxation time, while in our case the relaxation time is energy dependent. Ref. \cite{Pan} depicts that the second order correction to the particle distribution function $\partial f^{(2)}$ of Ref. \cite{Hamamoto} (written as an iterative solution to the Boltzmann transport equation within the
relaxation-time approximation)
does not satisfy the collision term of
the Boltzmann transport equation (BTE), and hence it is not self-consistent.  In Ref. \cite{Pan}, the derivation of $\partial f ^{(2)}$ considers the local change
in the equilibrium distribution function induced by the external fields and does not need to satisfy
the BTE, since the derivation is not associated with the evaluation of collision integral. In our study, we consider the approach of Ref. \cite{Pan} (see Eq. \ref{eqdfn}).

\section{Gapped 2D Rashba System}
\label{Sec3}
We consider a gapped two-dimensional electron gas (2DEG) with the Rashba spin-orbit 
interaction, where the Hamiltonian is given by
\begin{eqnarray}\label{ham-rash2d}
\label{eq10}
H&=&\frac{{ \hbar^2 \bf k^2}}{2m^{*}} \sigma_0 + 
\alpha \boldsymbol\sigma \cdot ({\bf k} \times {\bf \hat{z} }) + M \sigma_z.
\end{eqnarray} 
Here ${\bf k}=\{k\cos\phi,k\sin\phi\}$ is the electron's wavevector,
$\alpha$ denotes the Rashba spin-orbit interaction (RSOI) strength 
which measures the spin splitting induced by structural inversion asymmetry
and $M$ is the mass gap generated by breaking the time reversal symmetry.
The mass term $M$ can be generated either by applying an 
external magnetic field \cite{Culcer}
or by application of circularly polarized electromagnetic radiation \cite{Ojanen}.
 
Comparing Eq. (\ref{eq10}) with Eq. (\ref{eqH}), 
$d_x=\alpha k_y$, $d_y=-\alpha k_x$, $d_z=M$, 
$\phi^{\prime}=-(\pi/2 - \phi)$, 
$\theta^{\prime}=\tan^{-1}(s_{k}/c_{k})$, 
thus the energy spectrum is obtained as,
\begin{equation}
\label{eqd}
\epsilon_{\lambda}({\bf k}) = \frac{\hbar^2 k^2}{2m^{*}} + 
\lambda\sqrt{M^2+\alpha^2 k^2},
\end{equation}
and the corresponding normalized eigenstates are 
\begin{eqnarray}
\label{eqwf}
|{\bf k}, \lambda\rangle  = 
\sqrt{\frac{ 1+\lambda c_k}{2} } 
\left[\begin{array}{c}
1  \\
-i \frac{ \lambda s_k e^{i\phi}}{1+\lambda c_k} 
\end{array} \right],
\end{eqnarray}
where $c_{k}= M/\sqrt{M^2+\alpha^2 k^2}$ and 
$s_k = \alpha k/\sqrt{M^2+\alpha^2 k^2}$. 
There is a finite gap $2M$ at $k=0$  due to the time-reversal 
symmetry breaking term $M\sigma_z$.
The spin orientation of an electron with wavevector ${\bf k}$ in 
the gapped Rashba system is 
$\langle\boldsymbol\sigma\rangle_{\lambda} = 
\lambda\{s_{k}\sin\phi, -s_{k}\cos\phi, c_{k}\}$ and thus
the spin and linear momentum lock in such a way that 
$\langle\boldsymbol\sigma\rangle \cdot {\bf k}=0$.
Moreover, there is an out-of-plane spin orientation 
($\langle \sigma_z \rangle_{\lambda} = \lambda c_k$)
which is anti-parallel for the two bands and arises because of
the time-reversal symmetry breaking term.  
The Berry curvature corresponding to $\lambda$ band is given by
\begin{equation} 
{\boldsymbol \Omega}_\lambda({\bf k}) =  
-\lambda \frac{M  \alpha^2 {\bf \hat z} }
{2(M^2 + \alpha^2k^2)^{3/2} }.
\end{equation}
The isotropic Berry curvature is peaked at 
$ k = 0$ and decays with $ k$.

\begin{figure}
\includegraphics[width=82mm,height=60mm]{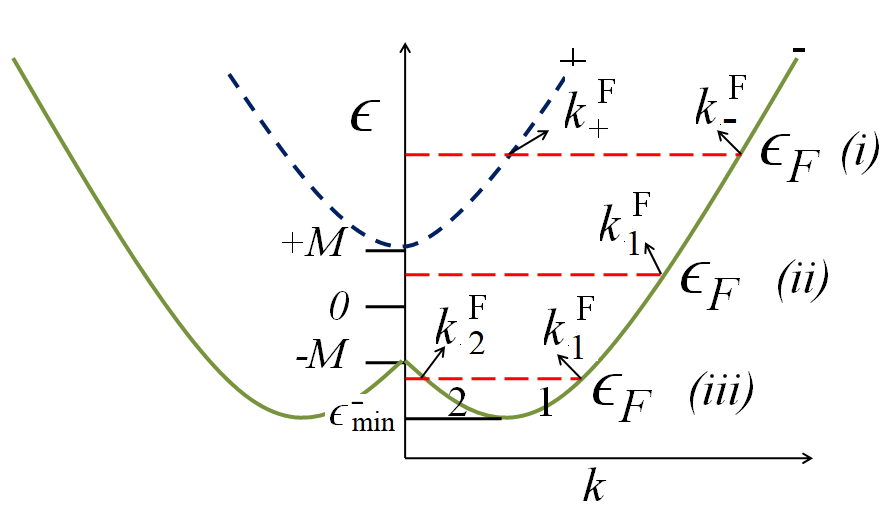}
\caption {Sketch of spin-split band structure of a  2D Rashba system with a Zeeman like
term ($M\sigma_z$), when $M <2 \epsilon_{\alpha}$.} 
%and (b) $M \geq 2\epsilon_{\alpha}$. Topology of the Fermi surfaces for
%different conditions are shown in (c) and (d).{\color{blue}FIGURE HAS TO BE MODIFIED}}
\label{Fig1}
\end{figure}

In Fig. (\ref{Fig1}), the band structure for the gapped Rashba systems given 
in Eq. (\ref{eqd}) is depicted with a fixed value of $\epsilon_\alpha$ for $M<2\epsilon_{\alpha}$.
The band $\epsilon_+({\bf k})$ attains a minimum energy 
$\epsilon_{\rm min}^{+} = +M$ at $k=0$ for all values of $M$. 
On the other hand,
the band $\epsilon_-({\bf k})$ attains a minimum energy 
$\epsilon_{\rm min}^{-} = -\epsilon_{\alpha}(1 + \tilde M^2) $  
at $k_m = k_\alpha \sqrt{1 - \tilde M^2}$, 
where $k_\alpha = m^*\alpha/\hbar^2$, 
$\epsilon_{\alpha} = m^*\alpha^2/(2\hbar^2)$ and 
$\tilde M = M/(2\epsilon_\alpha)$. 
It should be mentioned here that the above expression for 
$\epsilon_{\rm min}^{-}$ is valid only when $\tilde M < 1$.  
When $\tilde M \geq 1$,  
the minimum energy becomes $\epsilon_{\rm min}^{-} = -M$ at $k=0$.   
%It should be mentioned here that for $M=0$, the minimum energy is  
%$\epsilon_{\rm min}^{-} = - \epsilon_{\alpha}$ located at $k=k_{\alpha}$.
 
The wavevectors corresponding to $\epsilon>M$ (regime $(i)$, see Fig. \ref{Fig1}), are given by 
$k_{\lambda}=k_{\alpha} \sqrt{ (\tilde{E}-\lambda)^2-\tilde{M}^2} $,
where $\tilde{E} = \sqrt{1 + \tilde{\epsilon} + \tilde{M}^2}$ with 
$\tilde{\epsilon} = \epsilon/\epsilon_{\alpha}$.
Here, $k_{\pm}$ represent the radii of the two concentric circular constant energy surfaces.  For $\epsilon > M$, the topology of the Fermi surface has convex shape for both $\lambda = +$ and $\lambda =-$ bands. The density of states  
in each band is given by $D_{\lambda}(\epsilon)=D_0 (1 - \lambda/\tilde{E})$ 
with $D_0=m^*/2\pi\hbar^2$.

As mentioned earlier, for $\epsilon <- M$ (regime $(iii)$, see Fig. \ref{Fig1}), there exists only one energy band
$\epsilon_{-}$ and 
the topology of energy surface is completely different 
as compared to $\epsilon>M$. 
For $\epsilon< -M$ and $M<2\epsilon_{\alpha}$, the topology of the Fermi surface has concave-convex shape on
the inner and outer branches, respectively. For this regime, the wavevectors are represented by  $k_{\nu}=k_{\alpha} 
\sqrt{ [1+(-1)^{\nu-1}\tilde{E}]^2-\tilde{M}^2 }$ with $\nu=1,2$ ($\nu=1\rightarrow$ outer branch and $\nu=2\rightarrow$ inner branch). 
The DOS in each branch is given by 
$D_{\nu}=D_0 |1 + (-1)^{\nu-1}/\tilde{E} |$.

For the regime $-M\le\epsilon\le M$ (regime $(ii)$, see Fig. \ref{Fig1}), only 
$\nu=1$ branch exists with $\lambda=-1$.
Hence the DOS in this branch is given by
$D_{\nu=1}=D_0 (1 + 1/\tilde{E} )$.

The  generalized velocity components in the regime $(i)$ 
%($ \epsilon > M$)
are 
obtained as
\begin{align}
\label{eqv}
\langle \hat{v}_{x}\rangle_{\lambda}&=\frac{\hbar k_{\alpha}}{m^*}\tilde{E}\Big[1-\frac{\tilde{M}^2}{(\tilde{E}-\lambda)^2}\Big]^{1/2}\cos\phi+ \lambda\frac{\beta E_y \tilde{M}}{(\tilde{E}-\lambda)^3},\nonumber\\
\langle \hat{v}_{y}\rangle_{\lambda}&=\frac{\hbar k_{\alpha}}{m^*}\tilde{E}\Big[1-\frac{\tilde{M}^2}{(\tilde{E}-\lambda)^2}\Big]^{1/2}\sin\phi- \lambda\frac{ \beta E_x \tilde{M}}{(\tilde{E}-\lambda)^3},
\end{align}
where $\beta= e/(2\hbar k_{\alpha}^2)$. 
The velocity components in the regime $(iii)$
%($\epsilon_{\rm min}^{-} < \epsilon < - M$)
 with $M<2\epsilon_{\alpha}$
can be obtained from  Eq. (\ref{eqv}) with $\lambda  = -1$ 
and $\tilde{E}$ replaced by $(-1)^{\nu-1}\tilde{E}$.   
For the regime $(ii)$
%$ - M \le \epsilon \le M$, 
the $\langle \hat{v}_x\rangle$ and $\langle\hat{v}_y\rangle$ have the similar forms with $\nu=1$.

Similarly, the expectation values of spin velocity operators for the three regimes 
can be obtained. 
%\begin{figure*}[t]
%\begi}
%\begin{align}
%\label{eqsv}
%\langle v_{xx}\rangle_{\lambda}=&\frac{\hbar k_{\alpha}}{2m^*}\Big[\lambda\frac{(\tilde{E}-\lambda)^2-\tilde{M}^2}{\tilde{E}-\lambda}\sin2\phi\Big]
%+\frac{\beta E_y\tilde{M}[(\tilde{E}-\lambda)^2-\tilde{M}^2]^{1/2}}{(\tilde{E}-\lambda)^4}\sin\phi,\nonumber\\
%\langle v_{xy}\rangle_{\lambda}=&-\frac{\hbar k_{\alpha}}{m^*}\Big[1+\lambda\frac{(\tilde{E}-\lambda)^2-\tilde{M}^2}{\tilde{E}-\lambda}\cos^2\phi\Big]
%-\frac{\beta E_y\tilde{M}[(\tilde{E}-\lambda)^2-\tilde{M}^2]^{1/2}}{(\tilde{E}-\lambda)^4}\cos\phi,\nonumber\\
%\langle v_{yx}\rangle_{\lambda}=&\frac{\hbar k_{\alpha}}{m^*}\Big[1+\lambda\frac{(\tilde{E}-\lambda)^2-\tilde{M}^2}{\tilde{E}-\lambda}\sin^2\phi\Big]
%-\frac{\beta E_x\tilde{M}[(\tilde{E}-\lambda)^2-\tilde{M}^2]^{1/2}}{(\tilde{E}-\lambda)^4}\sin\phi,\nonumber\\
%\langle v_{yy}\rangle_{\lambda}=&-\frac{\hbar k_{\alpha}}{2m^*}\Big[\lambda\frac{(\tilde{E}-\lambda)^2-\tilde{M}^2}{\tilde{E}-\lambda}\sin2\phi\Big]
%+\frac{\beta E_x\tilde{M}[(\tilde{E}-\lambda)^2-\tilde{M}^2]^{1/2}}{(\tilde{E}-\lambda)^4}\cos\phi\nonumber\\
%\langle v_{xz}\rangle_{\lambda}=&\frac{\hbar k_{\alpha}}{m^*}\tilde{M}\Big[\lambda\frac{(\tilde{E}-\lambda)^2-\tilde{M}^2}{(\tilde{E}-\lambda)^2}\cos\phi\Big]
%+\frac{\beta E_y\tilde{M}^2}{(\tilde{E}-\lambda)^4},\nonumber\\
%\langle v_{yz}\rangle_{\lambda}=&\frac{\hbar k_{\alpha}}{m^*}\tilde{M}\Big[\lambda\frac{(\tilde{E}-\lambda)^2-\tilde{M}^2}{(\tilde{E}-\lambda)^2}\sin\phi\Big]
%-\frac{\beta E_y\tilde{M}^2}{(\tilde{E}-\lambda)^4}.
%\end{align}
%\end{widetext}  
%\end{figure*}
It is to be noted that the spin velocity  $\langle \hat{v}_{xz}\rangle$  and  $\langle \hat{v}_{yz} \rangle$ are zero for $M=0$. 
%For the regime $ \epsilon_{\rm min}^{-1} < \epsilon <-M$ with $M<2 \epsilon_\alpha$, 
%the expectation values of the spin velocity operators have the forms of 
%Eq. (\ref{eqsv}) with $\lambda \rightarrow-1$ and $\tilde{E}\rightarrow(-1)^{\nu-1}\tilde{E}$.
%For the regime $-M < \epsilon < M$, the expectation values of spin velocity operators take the similar forms with $\nu=1$. 

For calculating the second order spin currents, we need to know the relaxation time, which is calculated using the framework of semi-classical Boltzmann
transport equation including interband and intraband elastic scattering for regime $(i)$,
%$\epsilon> M$, 
and intrabanch and interbranch scattering within $\lambda=-1$ band for 
%$\epsilon_{\rm min}^{-} < \epsilon < - M$
regime $(iii)$  (see Appendix A).
The expressions for the relaxation time for the regime $(i)$
%$\epsilon> M$ 
are 
obtained as,
\begin{align}
\tau_{+}&=\frac{4\tau_0D_0}{A_{+}D_{+}+(B_++P_{+}/R)D_{-}},\nonumber\\
\tau_{-}&=\frac{4\tau_0D_0}{A_{-}D_{-}+(B_-+P_{-}R)D_{+}},
\end{align}
where $D_{\lambda}$ is the DOS,  
$\tau_0= 2\pi n_im V_0^2D_0/\hbar$,
$A_{\lambda}=1+3c_{k_\lambda}^2$, 
$B_{\lambda}=2(1-c_{k_\lambda}c_{k_{\lambda^{\prime}}})$, 
$P_{\lambda}=s_{k_\lambda}s_{k_{\lambda^{\prime}}}
v_{b}^{\lambda^{\prime}}/v_{b}^{\lambda}$, and
$R = [D_{-}(A_--P_+)+D_+B_-]/[D_{+}(A_+-P_-)+D_-B_+]$. 
Similarly, for the  regime $(iii)$
%$ \epsilon_{\rm min}^{-} < \epsilon < - M$,
the relaxation times are 
obtained as
\begin{align}
\tau_{1}&=\frac{4\tau_0D_0}{A_{1}D_{1}+(B_1-P_{1}/R)D_{2}},\nonumber\\
\tau_{2}&=\frac{4\tau_0D_0}{A_{2}D_{2}+(B_2-P_{2}R)D_{1}},
\end{align}
where $D_{\nu}$ is the DOS in each branch,
$A_{\nu}=1+3c_{k_\nu}^2$, 
$B_{\nu}=2(1+c_{k_\nu}c_{k_{\nu^{\prime}}})$,
$P_{\nu}=s_{k_\nu}s_{k_\nu^{\prime}} 
v_{b}^{ \nu^{\prime}}/v_{b}^{ \nu}$, and
$R = [D_{2}(A_2+P_1)+D_1B_2]/[D_{1}(A_1+P_2)+D_2B_1]$.
For the  regime $(ii)$,
% $-M \le \epsilon \le M$, 
only $\nu = 1$ branch persists. 
Thus, there exists only the intrabranch scattering and hence the relaxation time is 
$ \tau_- =  4\tau_0 D_0/D_1A_1$.
The analytical expressions for $\tau_{\lambda}$ and $\tau_{\nu}$ in terms of $\epsilon$ are cumbersome, thus not given here. 

\subsection{Background spin current}
\begin{figure}
\includegraphics[width=82mm,height=60mm]{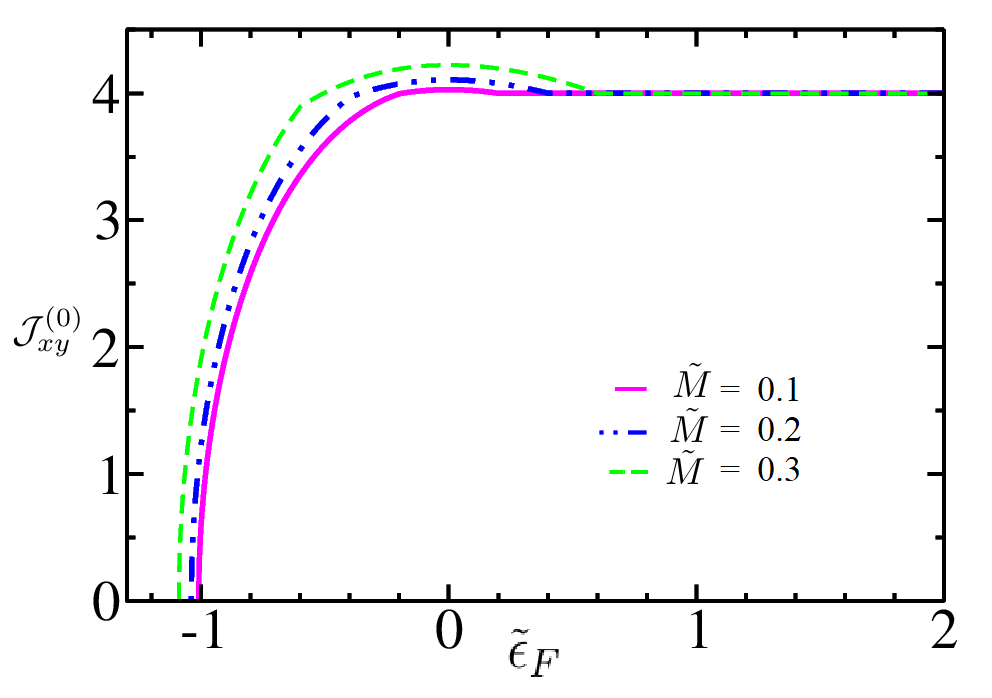}
\caption {The background spin current $\mathcal{J}_{xy}^{(0)}$ (in units of 
$\mathcal{J}_0$) as a function of 
 rescaled Fermi energy  $\tilde \epsilon_F$ for 
different values of  $\tilde{M}$.}
\label{BS}
\end{figure}

Fisrt we present the results for the non-propagating background spin current \cite{Rashba2}.  It can be easily shown that $\mathcal{J}_{xx}^{(0)}=\mathcal{J}_{yy}^{(0)}=0$. 
%{\color{blue}This is obvious because 2D Rashba spin-orbit interaction locks the spin orientation perpendicular to the electron momentum ($\langle\boldsymbol\sigma\rangle_{\lambda}\cdot {\bf k}=0$)} and 
%thereby yielding $\mathcal{J}_{xx}^{(0)}=\mathcal{J}_{yy}^{(0)}=0$.
Similarly, $\mathcal{J}_{xz}^{(0)}=\mathcal{J}_{yz}^{(0)} = 0$.
On the other hand, we obtain 
$\mathcal{J}_{xy}^{(0)} = - \mathcal{J}_{yx}^{(0)} \neq0$. 
At zero temperature $\mathcal{J}_{xy}^{(0)}$ [in units of 
$\mathcal{J}_0 = - \hbar^2 k_{\alpha}^3/(24 \pi m^*)$] has the following form 
\begin{eqnarray}
\mathcal{J}_{xy}^{(0)}
&=& 4, \;  \epsilon_F\geq M, 
\label{eq19a} \\\nonumber
\mathcal{J}_{xy}^{(0)} 
& = & (2 + 3 \tilde{E}_{F} - \tilde{E}_{F}^3 - 
2 \tilde{M}^3 \\\nonumber
&+& 3 \tilde{E}_{F} \tilde{M}^2 ), \; -M \leq\epsilon_F\leq M,
\label{eq19b}\\\nonumber
\mathcal{J}_{xy}^{(0)}
&=& 2 \tilde{E}_F (3 -  \tilde{E}_{F}^2 + 3 \tilde{M}^2), \;
\epsilon_F \leq -M, \label{eq19c}
\end{eqnarray}
where $\tilde{E}_{F} = \sqrt{1 + \tilde{\epsilon}_{F} + \tilde{M}^2}$ with $\tilde{\epsilon}_{F} = \epsilon_F/\epsilon_{\alpha}$. The background spin current $\mathcal{J}_{xy}^{(0)}$ (in units of 
$\mathcal{J}_0$) as a function of 
 rescaled Fermi energy  $\tilde \epsilon_F$ for  
different values of  $\tilde{M}$ is shown in Fig. (\ref{BS}).
When $\epsilon_F \geq M$ (or $\tilde{\epsilon}_F\geq 2\tilde{M}$, i.e. regime $(i)$), the background spin current is independent of 
$M$ and  $\epsilon_F$, whereas in the other two regimes, it 
depends on $M$ and $\epsilon_F$. 
Moreover, $\mathcal{J}_{xy}^{(0)}$ with $M \neq 0$ shows nonmonotonic
behavior for $ \epsilon_F < M$ (regime $(ii)$ and $(iii)$).
% first increases with decrease of $\epsilon_F$ and then starts decreasing 
%beyond the some charateristic value of $\epsilon_F$. 
The background spin current attains a maximum value 
$\mathcal{J}_{\rm max}^{(0)} 
= 2 \mathcal{J}_0[1 + (1+ \tilde M^2)^{3/2} - \tilde M^3] $ at $\epsilon_F = 0$
and vanishes at $\epsilon_F = \epsilon_{\rm min}^{-}$. 
The zeroth-order spin current is continuous,
while their first and second derivatives are discontinuous
at the band edges $\tilde{\epsilon}_F = \pm 2\tilde{M}$.
 
It is instructive to compare these results  with 
the results for 
$M=0$ case \cite{Rashba2}:
$\mathcal{J}_{xy}^{(0)}
=4 \mathcal{J}_0 $ for  $\epsilon_F \geq 0$ and  
$\mathcal{J}_{xy}^{(0)}
= 2\mathcal{J}_0 \sqrt{1 + \tilde{\epsilon}_F)} 
(2 - \tilde{\epsilon}_F)$ for  $\epsilon_F \leq 0$. 
These two equations and their first derivatives are continuous;
and the second derivative is discontinuous 
at $\epsilon_F = 0$.
 
%As mentioned earlier, though the zeroth-order spin current
%persists in the thermodynamic equilibrium, 
%it is is not associated with the any real spin transport and 
%cannot offer any spin injection or spin accumulation \cite{Rashba2}.

\subsection{Linear spin Hall current}
%According to symmetry analysis discussed in Sec. \ref{secc}, the linear current persists in a noncentrosymmetric system only when time reversal symmetry is broken. The Zeeman term in our system causes the time reversal symmetry breaking and produces a finite Berry curvature which should offer a linear spin current.  
Here, we present the results of linear spin current calculated using Eq. (\ref{eqja1}) for all possible combinations, where we find that
$\mathcal{J}_{xx}^{(1),\eta} = \mathcal{J}_{xy}^{(1),\eta} = \mathcal{J}_{yx}^{(1),\eta} =\mathcal{J}_{yy}^{(1),\eta}=0$ 
and $\mathcal{J}_{xz}^{(1),y} = - \mathcal{J}_{yz}^{(1),x}$. 
%(if $E_y$ in $\mathcal{J}_{xz,a}^{(1)}$ is equal to $E_x$ in $\mathcal{J}_{yz,a}^{(1)}$). 
The above results reveal that the electric fields 
cannot drive linear spin currents having in-plane spin polarization, 
whereas it can produce linear spin currents 
having out-of-plane spin polarization. The linear spin
current is always transverse to the electric field direction.

The expression for $\mathcal{J}_{xz}^{(1),y}$ ($=\mathcal{J}_{xz,a}^{(1),y}$) 
at zero temperature is obtained as
\begin{eqnarray}
\label{eq20a}
\mathcal{J}_{xz}^{(1),y}
& = & -\frac{ eE_0}{8\pi} \frac{(2\tilde{M}^2 - \tilde{M}^2 \tilde{\epsilon}_{F} - \tilde{\epsilon}_{F}^{2})}{(\tilde{\epsilon}_F + \tilde{M}^2)^2}, \;  \epsilon_F\geq M, \\\nonumber
\label{eq20b}
\mathcal{J}_{xz}^{(1),y}
& = & \frac{-eE_0}{16\pi}\Big[\frac{\tilde{M}^2}{(1+\tilde{E}_F)^2}-1\Big] \hspace{0.02in}\hspace{0.02in}, \;  -M \leq\epsilon_F\leq M, \\\nonumber
\label{eq20c}
\mathcal{J}_{xz}^{(1),y}
& = & \frac{eE_0}{4\pi} \frac{\tilde{M}^2 \tilde{E}_F}{(\tilde{\epsilon}_F + \tilde{M}^2)^2},  \;  \epsilon_F\leq -M.
\end{eqnarray}
%The above expressions are valid only for $M\neq0$, since for $M=0$, 
%the Berry curvature vanishes and no anomalous spin current exists.   
The linear spin current transverse to the electric field 
arising solely from the non-zero Berry curvature is the well 
known spin Hall current. 
This spin Hall current can be reversed by changing the electric field direction $\bf{E}$ to $-\bf{E}$. The spin Hall conductivity can be defined as
$ \sigma_s = \mathcal{J}_{xz,a}^{(1),y}/E_0$.
For $\epsilon_F \gg M$, the spin Hall conductivity is obtained as 
$\sigma_s^0 = e/8\pi$, which is exactly the same as
obtained by using the Kubo formula for $M\rightarrow0$ case by various groups 
\cite{Sinova,Moca} previously.  However, this universal result vanishes in presence of an arbitrary weak disorder \cite{Mishchenko}. Reference \cite{Dimitrova} also justifies this disappearance of static spin-Hall conductivity for any
non-vanishing disorder strength in case of the momentum-dependent Rashba strength and non-parabolic energy spectrum. Electron-electron interaction also modifies this universal value  of spin Hall conductivity \cite{Dimitrova}. Similar disorder effect can be performed on our calculation, which may provide the correction terms in the expressions of spin Hall conductivity.

\begin{figure}
\includegraphics[width=82mm,height=60mm]{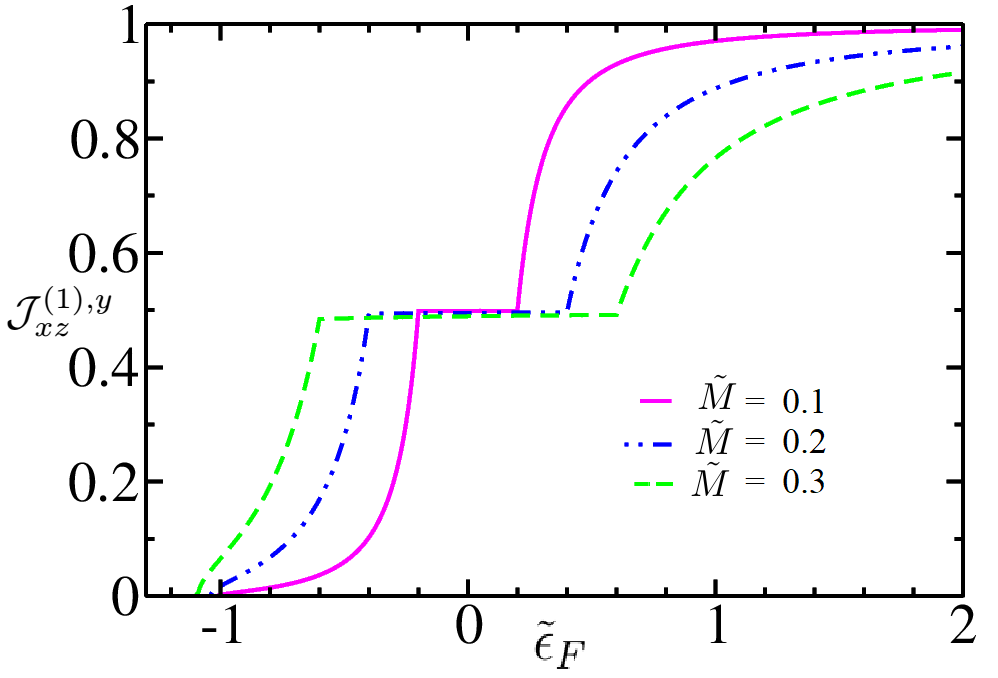}
\caption {The spin Hall current $\mathcal{J}_{xz}^{(1),y}$ (in units of $\mathcal{J}_{1}=\sigma_s^0E_0$)  
as a function of  rescaled Fermi energy $\tilde \epsilon_F$ for 
different values of $\tilde{M}$.}
\label{SHC}
\end{figure}

The linear spin Hall current $\mathcal{J}_{xz}^{(1),y}$ (in units of $\mathcal{J}_{1}=\sigma_s^0E_0$) 
as a function of  $\tilde{ \epsilon}_F$ for 
different values of $\tilde{M}$ is shown in Fig. (\ref{SHC}).
It is interesting to note that  the spin Hall current displays nearly
quantized plateau at $\mathcal{J}_{xz}^{(1),y} = (\mathcal{J}_{xz}^{(1),y})/2$ 
(i.e. half of the maximum value of spin Hall current) 
when Fermi energy lies between the two gap edges, i.e. 
$ -2\tilde{ M} < \tilde{\epsilon}_F <2\tilde{ M} $.
It reminds us the half-quantized anomalous charge Hall conductance
in gapped Rashba systems \cite{Culcer,Ojanen}.
Further, $\mathcal{J}_{xz}^{(1),y}$ decreases with increase of $\tilde{M}$ when 
$\tilde{\epsilon}_F >2\tilde{M}$, but
it increases with increase of $\tilde{M}$ when $ \tilde{\epsilon}_F <-2\tilde{M}$.

%{\bf the following paragraph can be given in the summary and conclusion section}:
%{\color{red}So far} we have explored the non-propagating spin current and linear spin Hall current. 
%As mentioned earlier, we have modeled our system in a way such that it can produce both the linear and nonlinear spin currents. 
%Before focusing on the results of non-linear spin current, mainly 
%quadratic spin currents, we concisely mention the major differences 
%between the spin Hall current (linear spin current) and the nonlinear spin current
%here. 
%(i) The origins of linear and nonlinear spin currents are {\color{red} uncorrelated}. 
%The nonlinear spin current discussed here is induced by the inversion symmetry breaking, while the origin of  spin Hall current is the finite Berry curvatures 
%of the carriers.  
%(ii) The nonlinear spin current is determined by the direction of electric field (elaborately discussed in Sec. \ref{Sec4c}), while spin Hall current is always perpendicular to the
%electric field. 

\subsection{Non-linear spin current}
\label{Sec4c}
In the previous sub-section, it is seen that a finite transverse linear spin
current exists while the longitudinal one vanishes.
% {\bf identically}???? for all values of Fermi energy.  
The first
non-vanishing contribution to longitudinal spin current in this system is quadratic in {\bf E}, as is the feature of IS-broken systems.  In this sub-section, the
quadratic spin current has been studied for all possible configurations of spin orientation, directions of charge propagation and applied electric field. 
First, we present results where only band velocity contributes to the quadratic spin current and
then we show that current with certain spin polarization arises only due to Berry curvature.
\\

{\bf Results for}  $\mathcal{J}_{xy}^{(2),\eta}$: 
First we consider the quadratic spin current $\mathcal{J}_{xy}^{(2),x}$, 
%having propagation in $x$ direction, spin polarization in $y$ direction,
%with applied electric field (bias) along $x$ direction(${\bf E} = E_x \hat{x}$)
so that only the band velocity contributes. It is to be noted that although a Hall field sets up along $\hat{y}$ direction due to anomalous drift of carriers, its contribution to nonlinear spin current ($\mathcal{J}_{xy}^{(2),y}$) is negligible as compared to that of the applied field, and hence not considered throughout the paper. 
%This current appears solely from the band velocity, as the anomalous component 
%of spin velocity (which is  $E_y$ dependent) is zero. 

\begin{figure}[h!]
\includegraphics[width=82mm,height=140mm]{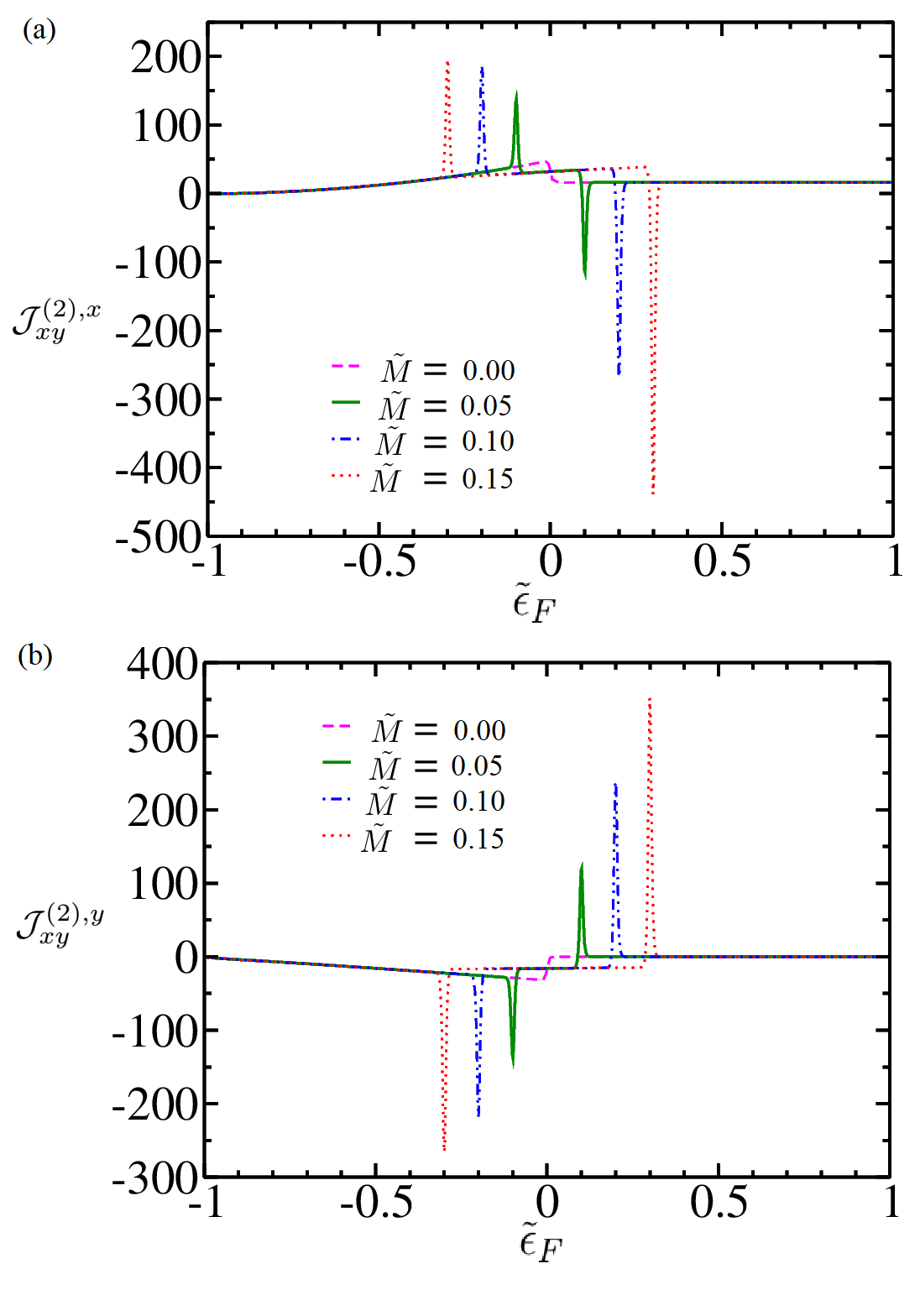}
\caption {The non-linear spin current (a) $\mathcal{J}_{xy}^{(2),x}$, and (b)$\mathcal{J}_{xy}^{(2),y}$   in
units of $\mathcal{J}_2$ as a function of $\tilde \epsilon_F $ for 
different values of $\tilde{M}$.} 
\label{NLS}
\end{figure}  
We present $\mathcal{J}_{xy}^{(2),x}$ as a function of 
$\tilde \epsilon_F$ for different values of $\tilde{M}$ in Fig. (\ref{NLS}a). 
%for ${\color{red}{\bf E} = E_x \hat{x}}$ .
The spin current is independent of the Fermi energy as well as $M$ when 
$\tilde{\epsilon}_F > 2\tilde{M}$. 
For the regime $-2\tilde{M}<\tilde{\epsilon}_F<2\tilde{M}$ and $\tilde{\epsilon}_F < -2\tilde{ M}$, the spin current is independent 
of $M$, but depends on the Fermi energy.
The spin current starts decreasing when $\tilde{\epsilon}_F < -2\tilde{ M}$.
There are two peaks with their values opposite 
in sign appearing at the gap edges, i.e, at $\tilde{\epsilon}_F=\pm 2\tilde{M}$. The absolute value of peak at $\tilde{\epsilon}_F = +2\tilde{ M}$ is greater than that of the $\tilde{\epsilon}_F = -2\tilde{M}$.  
Further, the peak value increases with the increasing strength of $\tilde{M}$.  
There is a sharp transition in the spin current around 
$ \epsilon_F = 0$ when $M=0$. 
%{\color{blue}The signature of nonlinearity appears through change in sign of the spin current??}. 
For $M=0$, using the formula in Eq. (\ref{eqjb22}), the spin current $\mathcal{J}_{xy}^{(2),x}$ at zero temperature is obtained as
\begin{align}
\label{J1}
\mathcal{J}_{xy}^{(2),x} & = 16 \mathcal{J}_2, \; \epsilon_F>0, \nonumber\\
\mathcal{J}_{xy}^{(2),x} & = \mathcal{J}_2 
\sqrt{1 + \tilde{\epsilon_F} }
\Big[48 + 72 \tilde{\epsilon}_F + 21 \tilde{\epsilon}_{F}^2
\Big], \; \epsilon_F<0,
\end{align} 
where $\mathcal{J}_2 =(e\tau_0 E_0/\hbar) ^2 \alpha/(32 \pi)$ with 
$\tau_0$ being unit of scattering time.

When electric field is directed along $\hat{y}$ direction (${\bf E}=E_y\hat{y}$),   the anomalous component ($\propto E_y$) of spin velocity, i.e, $v_{a,xy}$ exists.
However, this anomalous spin velocity gives no net contribution to non-linear spin current. %{\color{blue}because of the $\phi$ integration}. 
So, $\mathcal{J}_{xy}^{(2),y}$ appears from band component only.
The plots for $\mathcal{J}_{xy}^{(2),y}$ as a function of $\tilde \epsilon_F$ for different values of $\tilde{M}$ is shown in Fig. \ref{NLS}(b).   
%\begin{figure}[h!]
%\includegraphics[width=80mm,height=60mm]{J_xy_Ey.png}
%\caption {Plots of the spin current, $J_{xy}^{(2)}$ as a function of 
%$\tilde \epsilon_F$ for three different values of $M$. 
%Electric field is applied in $y$ direction. {\color{blue} I suggest all nonlinear plots to be %kept together in one figure, but in different frames. }}
%\label{Fig4}
%$\end{figure} 
Similar to the previous case, two peaks appear 
at $\tilde{\epsilon}_F=\pm 2\tilde{M}$.
Here, at $\tilde{\epsilon}_F=-2\tilde{M}$ the peak values are negative and positive at $\tilde{\epsilon}_F=+2\tilde{M}$, thereby following opposite trend of spin current when electric field is directed in $\hat{x}$ direction.
Thus, the polarization of these currents are opposite when driven by ${\bf E}=E_0\hat{x}$ and ${\bf E}=E_0\hat{y}$. 
For $M=0$, the analytical expressions of spin current
$\mathcal{J}_{xy}^{(2),y}$ at zero temperature  is obtained as
\begin{align}
\mathcal{J}_{xy}^{(2),y} & = 0, \; \epsilon_F>0,\nonumber\\
\mathcal{J}_{xy}^{(2),y} & = - \mathcal{J}_2 \sqrt{1 + \tilde{\epsilon}_F } \Big[32 +  16 \tilde{\epsilon}_F - 7 \tilde{\epsilon}_{F}^2 \Big], \; \epsilon_F<0.
\end{align} 
Thus, for the electric field in $\hat{y}$ direction, the spin current propagating in the $\hat{x}$ direction with the polarization in $\hat{y}$ direction is zero for  $\epsilon_F>0$, whereas it is non-zero for $\epsilon_F<0$.
When $\epsilon_F > 0$, the contribution from $\lambda = +$ and $\lambda=-$ bands 
are same in magnitude but opposite in sign, so their net contribution vanishes. %Further, to understand whether or not the energy dependent relaxation time has any role to reduce the value of $\mathcal{J}_{xy}^{(2)}$ to zero, we have calculated the spin currents for constant relaxation time.
However, it does not vanish
when the relaxation time is taken to be constant. \\
%So, its seems that the zero value for  $\mathcal{J}_{xy}^{(2)}$ for $\epsilon_F > 0 $ region is coming due to the energy dependent relaxation time $\tau(\epsilon)$.  

{\bf Results for} $\mathcal{J}_{yx}^{(2),\eta}$: 
\begin{figure}
\includegraphics[width=82mm,height=60mm]{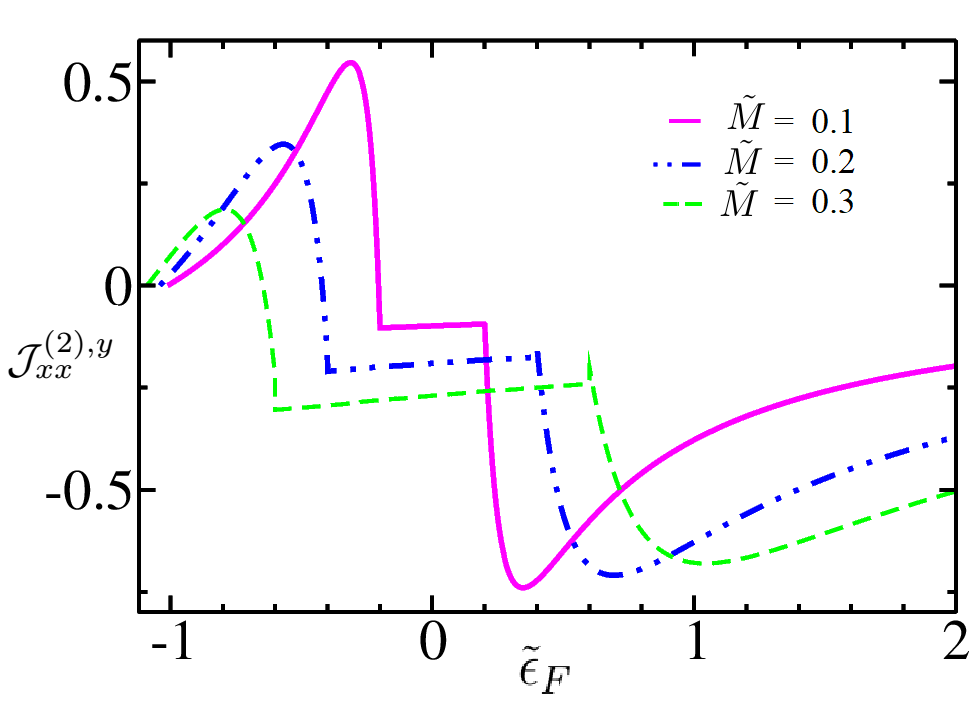}
  \caption {The spin current $\mathcal{J}_{xx}^{(2),y}$ (in units of $\mathcal{J}_2^{\prime}$) arising from the
anomalous velocity as a function of 
$\tilde \epsilon_F$ for different values of $\tilde{M}$. }
\label{Fig6}
\end{figure}  
Here we consider the opposite scenario, i.e, $\hat{y}$ directed spin current with polarization in $\hat{x}$ direction. 
As expected, we find $\mathcal{J}_{a,yx}^{(2),x}=0$ and
$\mathcal{J}_{b,yx}^{(2),x} \neq 0$ with
$\mathcal{J}_{b,yx}^{(2),x} = 
- \mathcal{J}_{b,xy}^{(2),y}$ and thereby $\mathcal{J}_{yx}^{(2),x} 
= - \mathcal{J}_{xy}^{(2),y}$. 
Similarly, we find 
$\mathcal{J}_{yx}^{(2),y}= 
- \mathcal{J}_{xy}^{(2),x}$. \\\\

{\bf Results for } $\mathcal{J}_{xx}^{(2),\eta} $ {\bf and}
$\mathcal{J}_{yy}^{(2),\eta} $: 
Now we shall present results for quadratic spin currents, 
$\mathcal{J}_{xx}^{(2),\eta}$ and $\mathcal{J}_{yy}^{(2),\eta} $, 
when both propagation and polarization are in the same direction.
For $\mathcal{J}_{xx}^{(2),x} $, 
the anomalous component of the spin velocity is zero and
contribution from the band velocity is also 
zero. 
Thus, we have $ \mathcal{J}_{xx}^{(2),x} = 0$.
Similarly, one can show that 
$ \mathcal{J}_{yy}^{(2),y}= 0$.

On the other hand, 
 we obtain that 
$\mathcal{J}_{b,xx}^{(2),y}$ is also zero,  
while %the nonlinear spin current due to the anomalous velocity
$\mathcal{J}_{a,xx}^{(2),y}$ would survive.
Using the similar analysis, we find that
$\mathcal{J}_{a,yy}^{(2),x}$ is finite.
Thus, one can generate pure anomalous nonlinear spin currents 
having propagation and polarization in the same direction,
while electric field is in their transverse direction.
Because of the isotropic nature of the Berry curvature, we find
$\mathcal{J}_{xx}^{(2),y}
= \mathcal{J}_{yy}^{(2),x} $.
The plot of $\mathcal{J}_{xx}^{(2),y}$ ($=\mathcal{J}_{a,xx}^{(2),y}$)  (in units of $\mathcal{J}_2^{\prime}$, where $\mathcal{J}_2^{\prime}=(\mathcal{J}_2 \hbar)/(\tau_0\alpha k_{\alpha}))$
as a function of $\tilde \epsilon_F$ is shown in Fig. (\ref{Fig6}). 
It displays that $\mathcal{J}_{xx}^{(2),y}$ is nearly flat when
$- 2\tilde{M} < \tilde{\epsilon_F} <2\tilde{ M}$. 
 The appearance of nonlinear anomalous spin current is reminiscent of Berry curvature induced nonlinear charge current which arises from the
dipole moment of the Berry curvature \cite{Sodemann,Nandy}. \\

%\begin{figure}[!htb]
 %   \centering
    %\begin{minipage}{.5\textwidth}
  %      \centering
   %     \includegraphics[width=0.8\linewidth, height=0.3\textheight]{J_NLS_A.png}
    %   \caption {The spin current $\mathcal{J}_{xx}^{(2),y}$ (in units of $\mathcal{J}_2^{\prime}$) arising from the
%anomalous velocity as a function of 
%$\tilde \epsilon_F$ for different values of $\tilde{M}$. }
 %       \label{Fig6}
   % \end{minipage}%
    %\begin{minipage}{0.5\textwidth}
  %      \centering
   %     \includegraphics[width=0.82\linewidth, height=0.3\textheight]{CC.png}
   %     \hspace{1cm}\caption {Comparison between the magnitudes of spin currents for different orders in a 2D gapped Rashba system.}
    %    \label{FigC}
   % \end{minipage}
%\end{figure}

{\bf Results for } $\mathcal{J}_{xz}^{(2),\eta} $ {\bf and}
$\mathcal{J}_{yz}^{(2),\eta} $: 
%We have seen that due to the non-zero Berry curvature, the transverse electric field can produce anomalous linear spin current (spin Hall current) having polarization ($\hat{z}$) perpendicular to direction of electric field and propagation. 
It is not possible to generate quadratic spin current having polarization in $\hat{z}$ direction, as we find $\mathcal{J}_{xz}^{(2),\eta}=\mathcal{J}_{yz}^{(2),\eta}=0$ .
%for any orientation of electric field for this system. }

All the above results of spin current (whether it is zero or non-zero) for a 2D gapped Rashba system are tabulated in Table I.

%{\color{blue} A TABLE MAY BE GOOD TO SHOW WHICH SPIN CURRENTS ARE ZERO/NON-ZERO FOR WHICH DIRECTIONS OF ELECTRIC FIELD.}
\begin{widetext}

\begin{table}
	
	\label{tab1}
		
		\begin{center}\begin{tabular}{ | c | c | c | c| c |c |} 
			\hline
			 Spin current & ${\bf E}=0$ & $\eta=x$ (B) & $\eta=x$ (A) & $\eta=y$ (B) & $\eta=y$ (A)\\
			\hline
			$\mathcal{J}_{xx}^{(0)} $ & 0 & NA & NA & NA & NA  \\
			\hline
			$\mathcal{J}_{xy}^{(0)} $ & Finite & NA & NA & NA & NA \\
			\hline
			$\mathcal{J}_{xz}^{(0)} $ & 0 & NA & NA & NA & NA	\\
			\hline
			$\mathcal{J}_{xx}^{(1),\eta} $ & NA & 0 &	0 & 0 &	0	\\
			\hline
			$\mathcal{J}_{xy}^{(1),\eta} $ & NA & 0 &	0 & 0 &	0   \\
			\hline
			$\mathcal{J}_{xz}^{(1),\eta} $ & NA & 0 & 0 & 0 & Finite  \\
			\hline
			$\mathcal{J}_{xx}^{(2),\eta} $ & NA & 0 & 0 & 0 & Finite \\
			\hline
			$\mathcal{J}_{xy}^{(2),\eta} $ & NA & Finite & 0  & Finite (for $\epsilon_F<0$) & 0	\\
			\hline
			$\mathcal{J}_{xz}^{(2),\eta} $ & NA & 0 & 0 & 0 & 0	\\
			\hline
		\end{tabular}\\
		\end{center}
		
		\hspace{1 cm}*NA: Not Applicable, *B: Band component contribution, *A: Anomalous component contribution
		\label{tab1}
		\caption{Nature of spin currents in 2D gapped Rashba system for different orientations of electric field ${\bf E}$. }
\end{table}
\end{widetext}

\begin{figure}
\includegraphics[width=85mm,height=60mm]{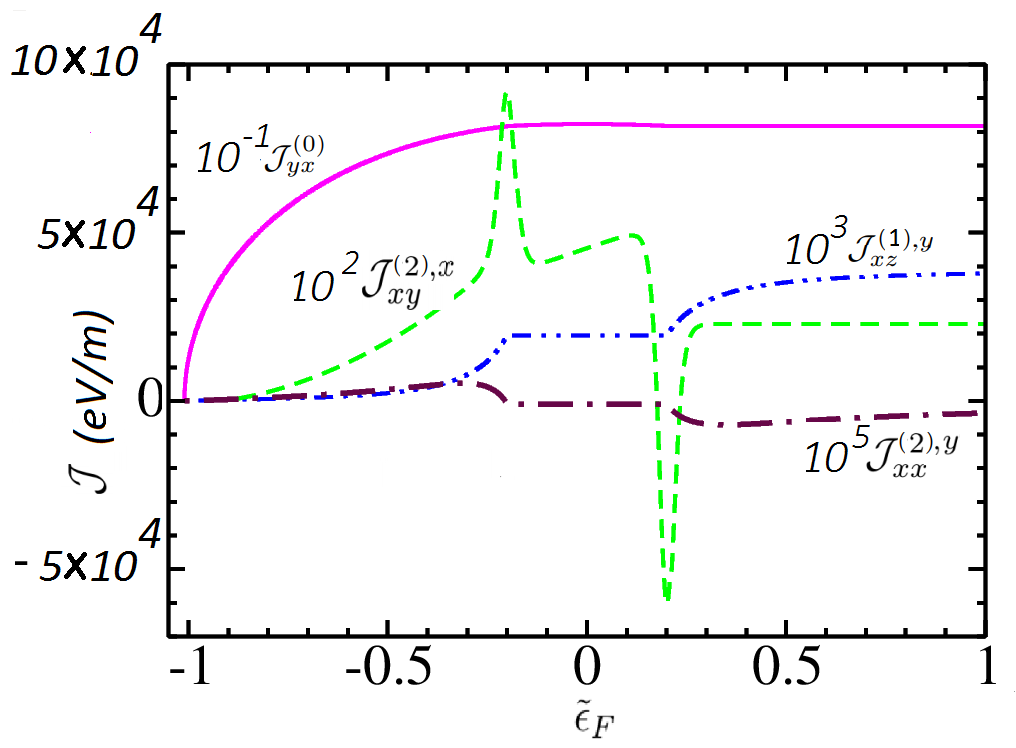}
  \caption {Comparison between the magnitudes of spin currents for different orders in a 2D gapped Rashba system.
      \label{FigC}}
\end{figure}

Here in Fig. \ref{FigC}, we compare the magnitudes of different orders of spin currents for a 2D gapped Rashba system with $E_0=10^3$ V/m, $\tau_0=2.5$ ps, $m^*=0.3m_e$ ($m_e$: electronic mass), $\alpha=0.1$ eV-nm, and $\tilde{M}=0.1$.
%Depending on the Rashba strength and the intensity of the applied electric field the ratio of different order spin currents drastically changes.  } \\

\section{3D noncentrosymmetric system}
\label{Sec4}
We consider 3D noncentrosymmetric metals such as 
Li${}_2$(Pd${}_{3-x}$Pt${}_x$)B and B20 compounds having cubic crystal structure. 
Based on symmetry analysis \cite{Kang,Samokhin,Lee}, 
the Hamiltonian for low-energy conduction electron is given by
\begin{equation} \label{3DH}
H = \frac{\hbar^2 {\bf k}^2}{2m^*} \sigma_0 + 
\alpha \boldsymbol{\sigma} \cdot {\bf k}.
\end{equation} 
Here ${\bf k} = k(\sin\theta \cos \phi, \sin \theta \sin \phi, \cos \theta) $ 
is the 3D Bloch wavevector and $\alpha$ is the strength of the
linear spin-orbit coupling. Comparing Eq. (\ref{3DH}) with Eq. (\ref{eqH}), we
have $ d_i = \alpha k_i$ with $i=x,y,z$ and  
the energy spectrum is given by
\begin{equation}
\epsilon_\lambda({\bf k}) = \frac{\hbar^2 k^2}{2m^*}  + \lambda \alpha k,
\end{equation}
where $\lambda = \pm$ denotes two chiral bands. The band 
$\epsilon_-({\bf k})$ has a minimum energy 
$\epsilon_{\rm min} = - \epsilon_\alpha $ at $k_\alpha$. 
Using the eigenstates given in  
Eq. (\ref{gen-wf}), the Berry curvature corresponding to $\lambda$ band is given as
\begin{equation}
{\boldsymbol \Omega}_\lambda = - \lambda \frac{ {\bf k} }{2 k^3}.
\end{equation}
The spin-momentum locking follows the constraint 
$ \langle {\boldsymbol \sigma} \rangle_\lambda \cdot {\bf k} = \lambda  k$
(or $ \langle {\boldsymbol \sigma} \rangle_\lambda \times {\bf k} =0$), 
which is completely opposite to the 2D case. Here we would like to mention that in 3D system, a gap in dispersion can't be created by adding Zeeman term and hence it is neglected in our calculation.

Using the Heisenberg's equation of motion, the band velocity operator is
given by $ {\bf \hat{v} }_b = \frac{\hbar {\bf k}}{m^*} \sigma_0
+ \frac{\alpha}{\hbar} {\boldsymbol \sigma}$. 
The band velocity expression for a given energy $\epsilon \ge 0$, 
%independent of bands $\lambda = \pm $,
is 
%\begin{equation}
${\bf v}_b = v_\alpha \sqrt{1 + \tilde \epsilon} \; \hat{\bf k}$
for both the bands $\lambda = \pm$.
Here $v_\alpha = \hbar k_\alpha/m^*$, 
$ \tilde \epsilon = \epsilon/\epsilon_\alpha$ and 
$ \hat {\bf k} = {\bf k}/k$ is the unit vector along the vector ${\bf k}$.
%\end{equation}
On contrary, for $\epsilon < 0$, the band velocity expression for
the two branches $\nu=1,2$ is
%\begin{equation}
${\bf v}_b = (-1)^{\nu -1} v_\alpha \sqrt{1 + \tilde \epsilon} \; 
{\bf \hat k}$.
%\end{equation}
%Here $v_\alpha = \hbar k_\alpha/m^*$, 
%$ \tilde \epsilon = \epsilon/\epsilon_\alpha$ and 
%$ \hat {\bf k} = {\bf k}/k$ is the unit vector along the vector ${\bf k}$.

For a given energy $\epsilon \ge 0$, the wavevector corresponding to band $\lambda$  is
$k_\lambda = k_\alpha [- \lambda  +  \sqrt{1 + \tilde \epsilon}]$ and
the density of states is given by
$$ 
D_\lambda(\epsilon) = D_3
\sqrt{\epsilon_\alpha} 
\Big[\frac{2+ \tilde \epsilon}{\sqrt{1 + \tilde \epsilon}}
- 2 \lambda \Big],
$$
where $ D_3=\frac{1}{4\pi^2}(\frac{2m^*}{\hbar^2})^{3/2}$.

On the other hand,
for a given energy $-\epsilon_\alpha <  \epsilon <0$, the 
wavevector corresponding to branch $\nu$  is
$k_\nu =  k_\alpha [1  + (-1)^{\nu+1} \sqrt{1 + \tilde \epsilon}]$ and
the density of states is given by
$$
D_\nu(\epsilon) = 
D_3
\sqrt{\epsilon_\alpha} 
\Big[ \frac{2+ \tilde \epsilon}{\sqrt{1 + \tilde \epsilon}}
+ 2 (-1)^{\nu+1} \Big].
$$
%({\bf we need to reqrite DOS expression much better way}.
The energy-dependent relaxation times used for calculating spin currents in 3D non-centrosymmetric metals have the following forms \cite{Sonu}
\begin{eqnarray}
\tau_{\lambda}&=&\frac{u_0}{2D_3}\Big[\frac{\sqrt{1+\tilde{\epsilon}}}{\sqrt{\epsilon_{\alpha}}(2+\tilde{\epsilon})}\Big]\Big[1-\lambda\frac{\sqrt{1+\tilde{\epsilon}}}{2+\tilde{\epsilon}}\Big], \; \epsilon\ge 0
	\\\nonumber
	\tau_{\nu}&=&\frac{u_0}{2D_3}\Big[\frac{\sqrt{1+\tilde{\epsilon}}}{\sqrt{\epsilon_{\alpha}}(2+\tilde{\epsilon})}\Big] \Big[1+(-1)^{\nu-1}\lambda\frac{\sqrt{1+\tilde{\epsilon}}}{2+\tilde{\epsilon}}\Big], \; \epsilon<0,
\end{eqnarray}
where $1/u_0=\pi n_{\rm imp} V_0^2/\hbar$ with $n_{\rm imp}$ being the impurity density.

\subsection{Background spin current}For a 3D Rashba system, the equilibrium background spin current 
$\mathcal{J}_{ii}^{(0)}$ with $i=x,y,z$
%  (= \mathcal{J}_{yy}^{(0)} = \mathcal{J}_{zz}^{(0)})$
is obtained as
\begin{equation}
\mathcal{J}_{ii}^{(0)} = \frac{\hbar^2 k_{\alpha}^4 }{3\pi^2 m^*} \sqrt{1 + 
\tilde \epsilon_F},\;
\forall \; \epsilon_F.
\end{equation}
Thus, the background spin current and its derivative are continuous across the BTP.
On the other hand, we find all other components are zero i.e. 
$\mathcal{J}_{ij}^{(0)} = 0 $ with $i \neq j$. 
%\mathcal{J}_{xz}^{(0)} = \mathcal{J}_{yx}^{(0)}
%= \mathcal{J}_{yz}^{(0)} = 0$.
The nature of background spin current in 3D is completely different from that 
of 2D Rashba system.
For $\tilde \epsilon_F \gg 1$, $  \mathcal{J}_{xx}^{(0)}  \sim \alpha^3$, but
$\mathcal{J}_{xx}^{(0)} \sim \alpha^4 $ when $ \tilde \epsilon_F \ll 1$.

\subsection{Spin Hall current} 
%Consider electric field is applied long $x$ direction,
Using symmetry analysis, the linear spin current due to the band
velocity becomes zero. On the other hand, the anomalous velocity 
gives rise to linear spin current which  propagates 
transverse to the external electric field.  
The spin Hall current expression is obtained as
$\mathcal{J}_{ij}^{(1),\nu} =\sigma_{\rm s}\epsilon_{ij\nu} E_{\nu}$, where the spin Hall conductivity $\sigma_{\rm s}$ is given by
\begin{equation}
\sigma_{\rm s} =
-\frac{e}{12 \pi^2} k_\alpha \sqrt{1 + \tilde \epsilon_F},\; \forall \; \epsilon_F. 
\end{equation}
Here $ \epsilon_{ij\nu}$ is the fully antisymmetric Levi-Civita tensor.
The same expression of $\sigma_{\rm s} $ can be obtained using 
the Kubo formula (see Appendix B).
%For other triad configurations of the electric field, propgation direction and
%spin polarization, we get
%$ \mathcal{J}_{yz}^{(1)}(E_0 \hat i) = \mathcal{J}_{yz}^{(1)}(E_0 \hat i) =
%\mathcal{J}_{yz}^{(1)}(E_0 \hat i) = $.
%The spin Hall conductivity is defined as 
%$ \sigma_{\rm sH} =   \mathcal{J}_{yz}^{(1)}/E_0$. 
For $ \tilde \epsilon_F \gg 1$, 
$\sigma_{\rm s} \simeq (-e /12\pi^2)\sqrt{2m^{*}\epsilon_F^0/\hbar^2}$ with
$\epsilon_F^0 $ being the Fermi energy for conventional 3D metals.
% which is nearly independent of the spin-orbit coupling strength $\alpha$.
On the other hand, $\sigma_{\rm s} \sim k_\alpha$ at the band touching point
${\bf k} = 0$.

\subsection{Nonlinear Spin current}
\begin{figure}
\begin{center}
\includegraphics[width=80mm,height=120mm]{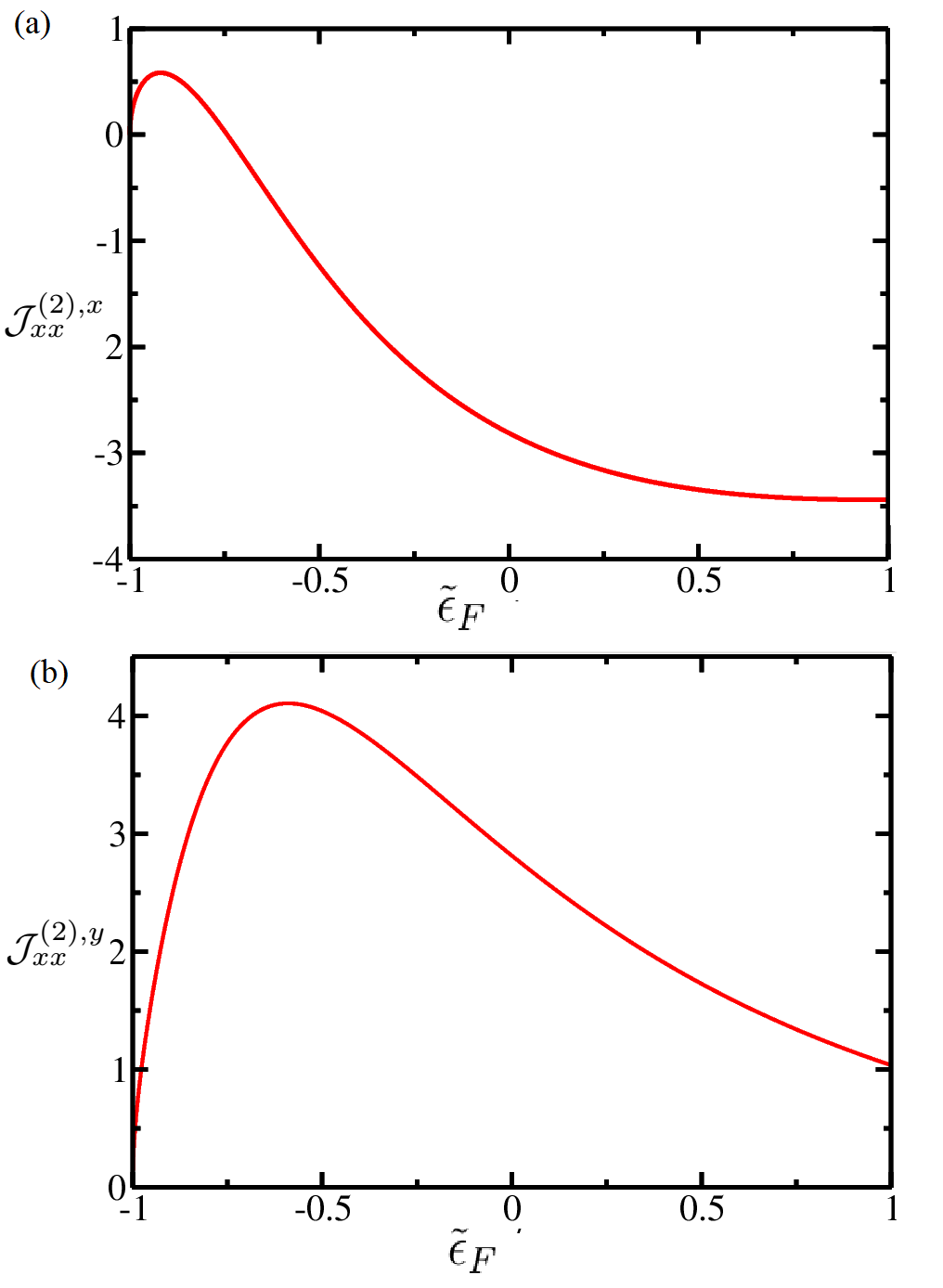}
\caption{Spin current (a) $\mathcal{J}_{xx}^{(2),x}$, and (b) $\mathcal{J}_{xx}^{(2),y}$ in units of $\mathcal{J}_3$ as a function of $\tilde{\epsilon}_F$. } %(b) Spin current, $J_x^x$ in each branch as a function of $\epsilon_F/\epsilon_{\alpha}$. Electric field in $x$ direction.}
\label{J3D}
\end{center}
\end{figure}  
Here we present the results of nonlinear spin currents for 3D Rashba system. The off diagonal spin current $\mathcal{J}_{ij}^{(2),\eta}$ with $i\neq j$ from both band and anomalous component of spin velocity are zero, %for any direction of electric field
whereas the diagonal spin currents $\mathcal{J}_{ii,b}^{(2),\eta}$ arising from band component have non zero values. 
 %This can be explained from {\color{blue}
%the fact that in 3D systems RSOI locks the spin parallel to electron momentum and hence the spin currents with propagation and polarization directions perpendicular to each other are zero.
The anomalous component of diagonal spin current, i.e, $\mathcal{J}_{ii,a}^{(2),\eta}$ is also zero. 
%for any orientation of electric field. 
 Thus, unlike the 
2D case (with Zemman term)  there is no nonlinear current in 3D due to the anomalous 
component because of the time reversal symmetry.\\\\
%Here we present the results of spin currents for 3D Rashba system when electric field is in $x$ direction. The off diagonal spin currents,  $J_x^y$, $J_x^z$, $J_y^x$, $J_y^z$, $J_z^x$ and $J_z^y$ vanish {\color{blue}due to the $\phi$ integration}, while the diagonal spin currents $J_x^x$, $J_y^y$  and $J_z^z$ have non zero values. This can be explained from {\color{blue}
%the fact that 3D RSOI locks the spin parallel to electron momentum.}
%Hence, the spin currents with propagation and polarization directions perpendicular to each other are zero.   
{\bf Results for $\mathcal{J}_{xx}^{(2),\eta}$}:
%Initially we consider that the electric field is in $\hat{x}$ direction.
As already mentioned for 3D Rashba system, the nonlinear spin current arises solely from the band component. The spin current $\mathcal{J}_{xx}^{(2),x}$ for a 3D system is obtained as
\begin{eqnarray}
\label{eq3d1}
\mathcal{J}_{xx}^{(2),x}&=&-\mathcal{J}_3\frac{\sqrt{1+\tilde{\epsilon}_F}}{(2+\tilde{\epsilon}_F)^5}\Big[90+239\tilde{\epsilon}_F\\\nonumber &+&199\tilde{\epsilon}_F^2 +58\tilde{\epsilon}_F^3+5\tilde{\epsilon}_F^4\Big],\;
 \forall \; \epsilon_F,
\end{eqnarray}
where $\mathcal{J}_3=(\hbar^2 eu_0 E_0\pi)^2/(30 m^{*3})$.
   
%{\color{red}For ${\bf E}=E_0 \hat{x}$, the spin current $J_x^x$ for 3D Rashba system is given as}
%\begin{eqnarray}
%\label{eq3d1}
%J_{x}^{x}&=-&\frac{\hbar(eu_0 E_x\pi)^2 }{30 m^{*3}}\frac{(1+\frac{\epsilon_F}{\epsilon_{\alpha}})^{\frac{1}{2}}}{(2+\frac{\epsilon_F}{\epsilon_{\alpha}})^5}\Big[90+ 239\frac{\epsilon_F}{\epsilon_{\alpha}}+ \\\nonumber &+&199\frac{\epsilon_F^2}{\epsilon_{\alpha}^2 } +58\frac{\epsilon_F^3}{\epsilon_{\alpha}^3}+5\frac{\epsilon_F^4}{\epsilon_{\alpha}^4}\Big],\hspace{0.2 in} {\color{red}\forall \; \epsilon_F}
%    for \hspace{0.05in}  both  \hspace{0.05in} \epsilon>0 \hspace{0.05in} and \hspace{0.05in} \epsilon<0 
%\end{eqnarray}

%Note that the spin current expressions obtained for the 3D case have {\color{red}identical} forms for  $\epsilon_F >0$ and $\epsilon_F<0$, while for 2D cases, the forms.
Expression (\ref{eq3d1}) depicts a smooth variation of quadratic spin current across $\epsilon_F=0$ (band touching point) in 3D Rashba, which ensures continuity of the first derivative of spin current, unlike its 2D counterpart.  This is because the forms of DOS are different for $\epsilon>0$ and $\epsilon<0$ in 2D Rashba, while for 3D case, they are same. 
The spin current, $\mathcal{J}_{xx}^{(2),x}$ (in units of $\mathcal{J}_3$) as a function  of $\tilde{\epsilon}_F$ is shown in Fig. \ref{J3D}(a) which depicts an increasing  trend  of spin currents (considering absolute value) with $\tilde{\epsilon}_F$, except in the region $-0.9< \tilde{\epsilon}_F<-0.75$.  The spin current changes sign at $\tilde{\epsilon}_F\sim 0.75$. 
%This can be {\color{red}explained from} the analytical expression of the spin current (see Eq. (\ref{eq3d1})). When $\tilde{\epsilon_F}<-0.75$, the negative contribution from {\color{red}second and fourth} terms (inside the square bracket) is {\color{green} more} than the positive contribution from the {\color{red}first, third and fifth terms. At $\tilde{\epsilon_F}=-0.75$, the contribution from those respective sets of terms cancel each other while for $\tilde{\epsilon_F}>-0.75$, the {\color{green}positive} contribution dominates over the {\color{green} negative} one.}

Now we consider the electric field in $\hat{y}$ direction. Similar to previous case (electric field in $\hat{x}$ direction), here also the spin current arises from band component only.  
The spin current $\mathcal{J}_{xx}^{(2),y}$ for a 3D system is obtained as
\begin{eqnarray}
\mathcal{J}_{xx}^{(2),y}&=&\mathcal{J}_3\frac{\sqrt{1+\tilde{\epsilon}_F}}{(2+\tilde{\epsilon}_F)^5}\Big[90+97\tilde{\epsilon}_F\\\nonumber
& +&2\tilde{\epsilon}_F^2-11\tilde{\epsilon}_F^3\Big], \; \forall \; \epsilon_F.     
\end{eqnarray}
The plot for  $\mathcal{J}_{xx}^{(2),y}$ as a function of $\tilde{\epsilon}_F$ is shown in Fig. \ref{J3D}(b). 
%\begin{figure}[h!]
%\label{Fig5}
%\begin{center}
%\includegraphics[width=80mm,height=80mm]{3D_yy.png}
%\includegraphics[width=80mm, height=80mm]{J_yy_Split.png}
%\caption {(a) Spin current, $J_{y/z}^{y/z}$ as a function of $\epsilon_F/\epsilon_{\alpha}$.} %(b) Spin current, $J_{y/z}^{y/z}$ in each branch as a function of $\epsilon_F/\epsilon_{\alpha}$. Electric field is in $x$ direction.}
%\end{center}
%\end{figure}  
The figure shows that the spin current $\mathcal{J}_{xx}^{(2),y}$ is zero at $\tilde{\epsilon}_F=-1$ and then it starts to increase, attends maxima at $\tilde{\epsilon}_F\sim-0.6$ and then again decreases. There is no sign change for the considered range of $\tilde{\epsilon}_F$. 
%{\color{blue}The reason behind non zero  spin current propagating in the perpendicular direction of the applied electric field is discussed earlier}.%The spin currents ($J_y/z^y/z$) coming from each branch are shown in Fig. (5b). Here also the major contribution is coming from $\nu=2$ (for $\epsilon_F<0$) and $\zeta=-$(for $\epsilon_F>0$) band.
% Further, the sign changing can be understood in other way also. In the expression the term $\frac{(1+\frac{\epsilon_F}{\epsilon_{\alpha}})^{\frac{1}{2}}}{(2+\frac{\epsilon_F}{\epsilon_{\alpha}})^5}$ will give the one of zeros of the spin current, i.e, $\epsilon_F=-\epsilon_{\alpha}$. %The term $\Big[90+ 239\frac{\epsilon_F}{\epsilon_{\alpha}}+199\frac{\epsilon_F^2}{\epsilon_{\alpha}^2 } +58\frac{\epsilon_F^3}{\epsilon_{\alpha}^3}+5\frac{\epsilon_F^4}{\epsilon_{\alpha}^4}\Big]$ will give the $4$ zeros of the equation, only one of them is $\epsilon_F\sim-0.75\epsilon_{\alpha}$, which is the real solution in the considered range of $\epsilon_F/\epsilon_{\alpha}$.

%The spin currents ($J_x^x$) coming from each branch are shown in Fig. (4b), where it is obtained that the major contribution is from $\nu=2$ (for $\epsilon_F<0$) and $\zeta=-$(for $\epsilon_F>0$) band.

Similarly, the spin currents $\mathcal{J}_{xx}^{(2),z}$ is obtained, where we find $\mathcal{J}_{xx}^{(2),z} =\mathcal{J}_{xx}^{(2),y}$.\\
%  which is expected by symmetry.
% and have the following form
%\begin{eqnarray}
%J_{y}^{y}= J_{z}^{z}&=&\frac{\hbar(eu_0 E_x\pi)^2 }{30 m^{*3}}\frac{(1+\frac{\epsilon_F}{\epsilon_{\alpha}})^{\frac{1}{2}}}{(2+\frac{\epsilon_F}{\epsilon_{\alpha}})^5}\Big[90+97\frac{\epsilon_F}{\epsilon_{\alpha}}\\\nonumber
%& +&2\frac{\epsilon_F^2}{\epsilon_{\alpha}^2 }-11\frac{\epsilon_F^3}{\epsilon_{\alpha}^3}\Big],\hspace{0.5 in}   {\color{red}\forall \; \epsilon_F } 
%for \hspace{0.1in}  both  \hspace{0.1in} \epsilon>0 \hspace{0.1in} and \hspace{0.1in} \epsilon<0 
%\end{eqnarray}

{\bf Results for $\mathcal{J}_{yy}^{(2),\eta}$ and $\mathcal{J}_{zz}^{(2)\eta}$}: Here we find $\mathcal{J}_{yy}^{(2),x}=\mathcal{J}_{yy}^{(2),z}=\mathcal{J}_{zz}^{(2),x}=\mathcal{J}_{zz}^{(2),y}=\mathcal{J}_{xx}^{(2),y}=\mathcal{J}_{xx}^{(2),z}$.
%which can be easily understood from symmetry argument.
Similarly, we obtain $\mathcal{J}_{yy}^{(2),y}=\mathcal{J}_{zz}^{(2),z}= \mathcal{J}_{xx}^{(2),x}$.\\

\begin{figure}
\includegraphics[width=82mm,height=60mm]{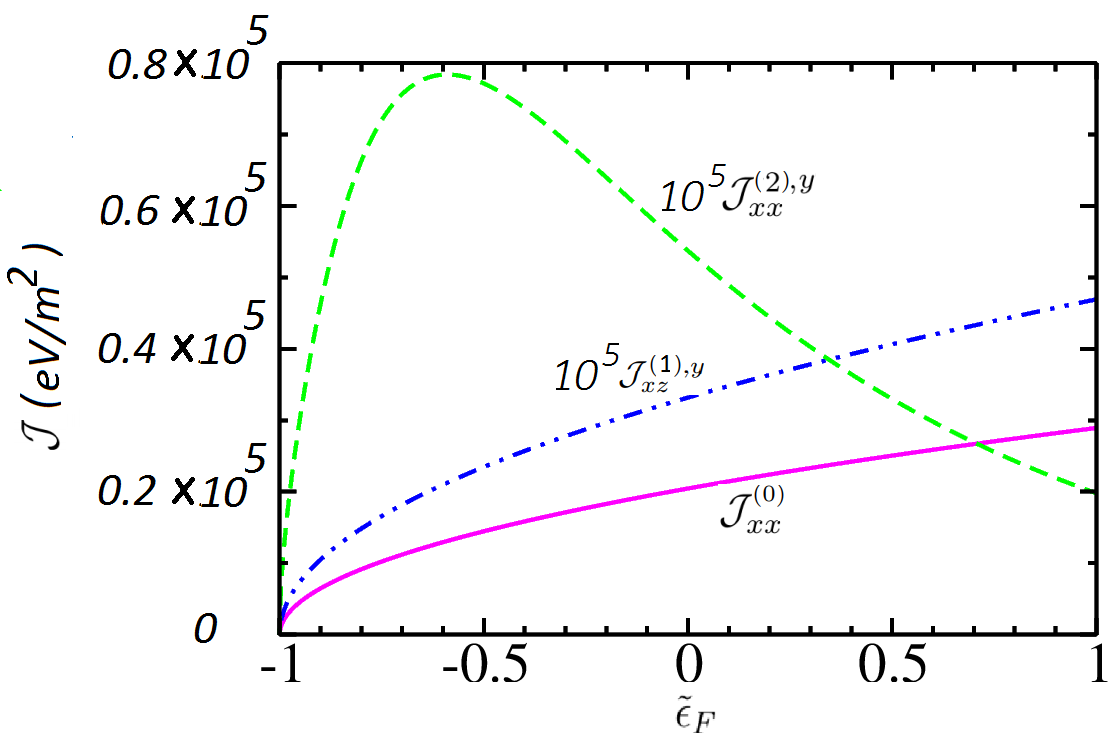}
  \caption {Comparison between the magnitudes of spin currents for different orders in a 3D  Rashba system.}
      \label{FigC2}
\end{figure}

 Fig. \ref{FigC2} depicts the comparison of the magnitudes of different orders of spin currents for a 3D Rashba system with $E_0=10^3$ V/m, $\tau_0=u_0/(2D_3\sqrt{\epsilon_{\alpha}})=2.5$ ps, $m^*=0.3m_e$ ($m_e$: electronic mass), and $\alpha=0.1$ eV-nm.\\
%Depending on the Rashba strength and the intensity of the applied electric field the ratio of different order spin currents drastically changes.  } \\

The results of spin currents in 3D noncentrosymmetric system are concisely summarized in Table. II.

\begin{widetext}

\begin{table}
	
	\label{tab1}
		
		\begin{center}\begin{tabular}{ | c | c | c | c | c | c | c | c |}
			\hline
			Spin current & ${\bf E}=0$ & $\eta=x$ (B)  & $\eta=x$ (A) & $\eta=y$ (B) & $\eta=y $ (A) & $\eta=z$ (B) & $\eta=z$ (A) \\
			\hline
			$\mathcal{J}_{xx}^{(0)} $ & Finite & NA & NA  & NA   & NA & NA  & NA\\
			\hline
			$\mathcal{J}_{xy}^{(0)} $ & 0 & NA & NA  & NA & NA & NA  & NA\\
			\hline
			$\mathcal{J}_{xz}^{(0)} $ & 0 & NA & NA  & NA & NA & NA  & NA	\\
			\hline
			$\mathcal{J}_{xx}^{(1),\eta} $ & NA & 0 &	0 & 0 & 0 &	0 & 0	\\
			\hline
			$\mathcal{J}_{xy}^{(1),\eta} $ & NA & 0 &	0 & 0 & 0 & 0 & Finite  \\
			\hline
			$\mathcal{J}_{xz}^{(1),\eta} $ & NA & 0 & 0 & 0  & Finite & 0 & 0\\
			\hline
			$\mathcal{J}_{xx}^{(2),\eta} $ & NA & Finite & 0  & Finite & 0 & Finite & 0\\
			\hline
			$\mathcal{J}_{xy}^{(2),\eta} $ & NA & 0 & 0 & 0  & 0 &	0 & 0	\\
			\hline
		    $\mathcal{J}_{xz}^{(2),\eta} $ & NA & 0 & 0 & 0  & 0 &	0 & 0	\\
			\hline
		\end{tabular}\\
		\end{center}
		
		\hspace{1 cm}*NA: Not Applicable, *B: Band component contribution, *A: Anomalous component contribution
		\label{tab1}
		\caption{Nature of spin currents in 3D noncentrosymmetric system for different orientations of electric field ${\bf E}$. }
\end{table}
\end{widetext}

Here we would like to emphasize on the differences of the 2D and 3D results.
In case of 2D Rashba system, the spin currents arising from band component always have
the polarization and propagation direction perpendicular to each other, whereas in 3D system, the propagation and polarization directions are same. This may be attributed to fact that for 2D case,  Rashba
spin-orbit interaction ($\alpha \boldsymbol\sigma \cdot ({\bf k} \times {\bf \hat{z} }$)) locks the spin orientation perpendicular to the momentum and thereby yields no contribution for the spin currents having same propagation and polarization directions. For 3D case,  Rashba
spin-orbit interaction ($\alpha \boldsymbol{\sigma} \cdot {\bf k}$) locks the spin orientation parallel to the momentum and hence the spin currents having the propagation and polarization directions perpendicular to each other become zero.  For 3D case, the expressions for the spin currents depicts a smooth variation across $\epsilon_F=0$,  unlike the 2D case. This
is because of the different forms of DOS for $\epsilon_F > 0$ and
$\epsilon_F < 0$ in 2D Rashba, while for 3D case, they are same.  For 2D  gapped Rashba system, the Berry curvature induces pure nonlinear spin current, whereas in 3D system, there is no anomalous nonlinear spin current because of the time reversal symmetry.

\section{Conclusion}
\label{Sec5}
We have studied the background, linear and nonlinear spin currents in 2D Rashba spin-orbit coupled systems with Zeeman coupling and in 3D non-centrosymmetric metals. We have incorporated the correction due to Berry curvature induced anomalous velocity in the semiclassical equations of motion, which contributes to spin currents transverse to the applied field. We have considered energy dependent relaxation time obtained by solving the Boltzmann transport equations self consistently with (i) interband and intraband scattering for $\epsilon> M$, (ii) intrabanch scattering for $-M\le\epsilon\le M$ and (iii) interbranch and intrabranch scattering for $\epsilon<-M$  in presence of the short-range impurity. 

For 2D Rashba systems, the background spin current has only an in-plane component with spin polarization perpendicular to direction of propagation. It increases with $\epsilon_F$ and attains a fixed value (independent of the Zeeman coupling) when Fermi energy is above the `Zeeman' gap.  The linear spin current has only a transverse component due to anomalous velocity of carriers. The spin  Hall conductivity rises with $\epsilon_F$, exhibits a plateau at the Zeeman gap (similar to the Hall plateau) and saturates to the intrinsic value $e/(8\pi)$ at higher $\epsilon_F$, which is independent of the Zeeman gap. This linear current has out-of-plane spin polarization. The nonlinear spin current (arising from band component) has both longitudinal and transverse components in general and are polarized in-plane. When $\epsilon_F$ is above the gap, the longitudinal part is constant in $\epsilon_F$ while the transverse one vanishes. Both the nonlinear components are sharply peaked at the gap edges with opposite spin polarizations at the upper and lower edges. The magnitudes of peak values get enhanced with the strength of the Zeeman coupling. We get pure anomalous nonlinear spin current with polarization along the direction of propagation with extrema near the gap edges.

In 3D noncentrosymmetric metals, the background spin current has spin polarization along the direction of propagation and is an increasing function of $\epsilon_F$. The linear spin current has its directions of propagation, spin polarization and applied electric field mutually orthogonal to one another and  increases as a function of $\epsilon_F$.   For very high $\epsilon_F$, the linear spin Hall conductance is nearly independent of Rashba coupling strength and varies as $\sqrt{\epsilon_F^0}$. Both the transverse and longitudinal components of nonlinear spin current have their spin polarization aligned parallel or anti-parallel to the direction of propagation.\\

Thus, gapped 2D Rashba systems and 3D noncentrosymmetric metals are
valuable assets to explore the Berry curvature induced spin currents. The
correction due to Berry curvature in the spin velocity operator results in
an `extrinsic' spin Hall current and gives an additional $xx$ (or $yy$) component of
nonlinear spin current.  The magnitudes of these currents in gapped
Rashba systems can be controlled by tuning the external magnetic field,
which makes it suitable for experimental studies. The plateau of linear
spin current and the sharp peaks of nonlinear spin currents may act as
probe for detection of Zeeman coupling from magnetic impurities and its
corresponding strength in 2D Rashba systems. In 3D noncentrosymmetric
metals, the Berry curvature results in linear spin Hall current but does
not affect the nonlinear ones.

\section{Acknowledgments}
P. Kapri thanks Department of Physics, IIT Kanpur, India for financial support.

\appendix
\begin{widetext}
\section{Derivation of Relaxation time for a 2D gapped Rashba system }
Here we present the derivation of relaxation time of a gapped 2D Rashba system 
with spin-independent short-range scatterer using the semiclassical BTE 
self-consistently. 
Following Refs. \cite{Sonu,Ma},  the coupled equations for the relaxation time $\tau_{\zeta} (\epsilon)$ are given by 
\begin{equation} \label{coupled}
\frac{1}{\tau_{\zeta}(\epsilon)}=\frac{1}{2\pi}\sum_{\zeta^{\prime}}\int d\epsilon_{\zeta^{\prime}}d\phi^{\prime}
D_{\zeta^{\prime}}(\epsilon_{\zeta^{\prime}})W_{\zeta^{\prime}\zeta}(\epsilon_{\zeta},\epsilon_{\zeta^{\prime}})\Big[1-\cos(\phi^{\prime}-\phi)
\frac{v_b^{\zeta^{\prime}}\tau_{\zeta^\prime}}{v_{b}^{\zeta}\tau_{\zeta}}\Big].
\end{equation}
Here $\zeta \equiv (\lambda, {\bf k})$ and $\zeta \equiv(\nu, {\bf k})$ is the eigenstate index for the regime $\epsilon > M$ and 
$\epsilon_{\rm min}^{-} < \epsilon < -M$ respectively
and the transition rate between the states $\zeta$ and $\zeta^{\prime}$ is
\begin{align}
W_{\zeta^{\prime},\zeta} & = \frac{2\pi}{\hbar}
\Big|\langle \zeta^{\prime}|V({\bf r})| \zeta\rangle  \Big|^2 
\delta(\epsilon_{\zeta} - \epsilon_{\zeta^{\prime}}),\nonumber
\end{align}
where $V({\bf r}) = V_0 \sum_i \delta({\bf r} - {\bf R}_i)$ with a 
constant strength $V_0$.
% $\tau_0=\frac{2\pi n_im V_0^2D_0}{\hbar}$ with $n_{im}$ being the impurity %concentration and $D_0=\frac{m^*}{2\pi\hbar^2}$.

After performing the integral and summation of Eq. (\ref{coupled}), 
for the regime $\epsilon> M$, it reduces  to
%\begin{equation}
%\frac{1}{\tau_{\lambda}}=\frac{D_{\lambda}}{4\tau_0D_0}\Big[(1+\lambda c_{\lambda})^2+\frac{s_{\lambda}^4}{(1+\lambda c_{\lambda})^2}-s_{\lambda}^2\Big]+\frac{D_{\lambda^{\prime}}}{4\tau_0D_0}\Big[(1+\lambda c_{\lambda})(1+\lambda^{\prime} c_{\lambda^{\prime}})+\frac{s_{\lambda}^2s_{\lambda^{\prime}}^2}{(1+\lambda c_{\lambda})(1+\lambda^{\prime} c_{\lambda^{\prime}})}+s_{\lambda}s_{\lambda^{\prime}}\frac{v_b^{\lambda^{\prime}}\tau_{\lambda^{\prime}}}{v_b^{\lambda}\tau_{\lambda}}\Big],
%\end{equation}
%where $\lambda \neq \lambda^{\prime}$, 
%$\tau_0=\frac{2\pi n_{\rm im} V_0^2 D_0}{\hbar}$ with $n_{\rm im}$ being 
%the impurity concentration and $D_0 = m^*/(2\pi\hbar^2)$.
%With the consideration $A_{\lambda}=(1+\lambda c_{\lambda})^2+\frac{s_{\lambda}^4}{(1+\lambda c_{\lambda})^2}-s_{\lambda}^2=1+3c_{\lambda}^2$, $B_{\lambda}=(1+\lambda c_{\lambda})(1+\lambda^{\prime} c_{\lambda^{\prime}})+\frac{s_{\lambda}^2s_{\lambda^{\prime}}^2}{(1+\lambda c_{\lambda})(1+\lambda^{\prime} c_{\lambda^{\prime}})}=2(1-c_{\lambda}c_{\lambda^{\prime}})=B_{\lambda^{\prime}}$, and $P_{\lambda}=s_{\lambda}s_{\lambda^{\prime}}\frac{v_b^{\lambda^{\prime}}}{v_b^{\lambda}}$,
\begin{equation}
\frac{1}{\tau_{\lambda}}=\frac{D_{\lambda}A_{\lambda}}{4\tau_0D_0}+\frac{D_{\lambda^{\prime}}(B_{\lambda}+(P_{\lambda}\tau_{\lambda^{\prime}})/\tau_{\lambda})}{4\tau_0D_0},
\end{equation}
where $A_{\lambda} = 1 + 3 c_{k_\lambda}^2$, 
$B_{\lambda}= 2(1-c_{k_\lambda} c_{k_{\lambda^{\prime}}}) = B_{\lambda^{\prime}}$, and $P_{\lambda} = s_{k_\lambda} s_{k_{\lambda^{\prime}}} v_b^{\lambda^{\prime}}/v_b^{\lambda}$. Also, 
$\tau_0 = 2\pi n_{\rm im} V_0^2 D_0/\hbar$ with $n_{\rm im}$ being 
the impurity concentration and $D_0 = m^*/(2\pi\hbar^2)$.
Solving the coupled equations for $1/\tau_+$ and $1/\tau_-$, the relaxation times of the two bands for $\epsilon> M$ are obtained as
\begin{align}
\tau_{+}=\frac{4\tau_0D_0}{A_{+}D_{+}+(B_++P_{+}/R)D_{-}}\hspace{0.1in};\hspace{0.1in}
\tau_{-}=\frac{4\tau_0D_0}{A_{-}D_{-}+(B_-+P_{-}R)D_{+}},
\end{align}
with $R=\frac{D_{-}(A_--P_+)+D_+B_-}{D_{+}(A_+-P_-)+D_-B_+}$.

Similarly, for the regime $\epsilon_{\rm min}^{-} < \epsilon < -M$, 
the coupled equations for the relaxation time $\tau_{\nu}(\epsilon)$ are
\begin{equation}
\label{eqA10}
\frac{1}{\tau_{\nu}}=\frac{D_{\nu}A_{\nu}}{4\tau_0D_0}+\frac{D_{\nu^{\prime}}(B_{\nu} - (P_{\nu}\tau_{\nu^{\prime}})/\tau_{\nu})}{4\tau_0D_0}.
\end{equation}
Here $A_{\nu} = 1+3c_{k_\nu}^2$, $B_{\nu} = 2(1+c_{k_\nu}c_{k_{\nu^{\prime}}})=B_{\nu^{\prime}}$, and $P_{\nu}=s_{k_\nu}s_{k_{\nu^{\prime}}} v_b^{\nu^{\prime}}/v_b^{\nu}$.
Solving the Eq. \ref{eqA10}, the relaxation times of the two branches for $ \epsilon_{\rm min}^{-} < \epsilon < - M$ are obtained as
\begin{align}
\tau_{1}=\frac{4\tau_0D_0}{A_{1}D_{1}+(B_1-P_{1}/R)D_{2}}\hspace{0.1in};\hspace{0.1in}
\tau_{2}=\frac{4\tau_0D_0}{A_{2}D_{2}+(B_2-P_{2}R)D_{1}},
\end{align}
where $R=\frac{D_{2}(A_2 + P_1) + D_1 B_2}{D_{1}(A_1 + P_2) + D_2 B_1}$.

\section{Spin Hall current in  3D noncentrosymmetric system from Kubo formalism}
The spin Hall conductivity in a clean 3D Rashba system using Kubo formula \cite{Moca} is obtained as 
\begin{eqnarray}
\sigma_{\rm s}&=&\frac{e\hbar^2}{2}\sum_{\lambda\neq \lambda^{\prime}}\int\frac{d^3\bf{k}}{(2\pi)^3}
(f_{k_{\lambda^{\prime}}}-f_{k_{\lambda}})
 Im\frac{\langle k, \lambda^{\prime}|\hat{v}_{xz}| k, \lambda\rangle
\langle k, \lambda^{\prime}|\hat{v}_{y}| k, \lambda\rangle}{(\epsilon_{k_{\lambda}}-\epsilon_{k_{ \lambda^{\prime}}})^2}\\\nonumber
&=&\frac{e\hbar^2}{4m^{*}\alpha}\int \frac{d^3\bf{k}}{(2\pi)^3}(f_--f_+)\frac{\sin^2\theta\cos^2\phi}{k}
=\frac{ek_{\alpha}}{12\pi^2}\sqrt{1+\tilde{\epsilon}_F}.
\end{eqnarray}

\end{widetext}

\end{document}